\documentclass[
    aps,      
    prx,      
    reprint,  
    noeprint, 
    nofootinbib, 
]{revtex4-2}
\usepackage{
    physics,
    mathtools,
    amssymb,
    bm,
    hyperref,
    xcolor,
    graphicx,
}
\graphicspath{{./graphics/drafts2/}{./graphics/drafts1/}}
\usepackage[version=4]{mhchem}
\usepackage{siunitx}
\DeclareSIUnit\angstrom{\text {Å}}
\usepackage{threeparttable}

\DeclareMathOperator{\sign}{sign}
\usepackage[capitalize]{cleveref}
\usepackage[inline]{enumitem}

\usepackage{booktabs}
\usepackage{csvsimple}

\usepackage{multirow}
\newcommand{\RNum}[1]{\uppercase\expandafter{\romannumeral #1\relax}}

\begin{document}
\title{Progress in superconductor--semiconductor topological {Josephson} junctions}
\author{William F. Schiela}
\email{william.schiela@nyu.edu}
\author{Peng Yu}
\email{py2071@nyu.edu}
\author{Javad Shabani}
\email{jshabani@nyu.edu}
\affiliation{Center for Quantum Information Physics, New York University, New York, NY 10003, USA}
\date{\today}

\begin{abstract}

Majorana bound states (MBSs) are quasiparticles which are their own antiparticles.  They are predicted to emerge as zero-energy modes localized at the boundary of a topological superconductor.  No intrinsic topological superconductor is known to date. However, by interfacing conventional superconductors and semiconductors with strong spin--orbit coupling it is possible to create a system hosting topological states. Hence epitaxial superconductors and semiconductors have emerged as an attractive materials system with atomically sharp interfaces and broad flexibility in device fabrications incorporating Josephson junctions. We discuss the basics of topological superconductivity and provide insight on how to go beyond current state-of-the-art experiments. We argue that the ultimate success in realizing MBS physics requires the observation of non-Abelian braiding and fusion experiments.





\end{abstract}
\maketitle


\section{Introduction}

Topological superconductors have proven far more elusive than other topological condensed matter phases, but they continue to garner interest both for their unusual physical properties and their potential practical applications in quantum information science. From the perspective of fundamental physics, topological superconductors host quasiparticle excitations that form at their boundaries in both one and two dimensions. These quasiparticles satisfy the defining property of Majorana fermions \citep{majorana1937,*majorana1937_english,kitaev2001_unpairedMajoranas} in that their creation and annihilation operators are identical and self-adjoint; in other words, they are their own antiparticles. No such fermionic elementary particle has yet been found in nature (though a case has been made for neutrinos \citep{elliott&franz2015colloquium,avignone2008}). In the present context, however, ``fermion'' is a misnomer: though they are composed of fermions, Majorana bound states (MBSs) in reduced spatial dimensions do not obey fermionic particle statistics. Instead, they exhibit non-Abelian statistics \citep{moore&read1991,read&green2000,ivanov2001} whereby exchanging two MBSs can realize a nontrivial unitary transformation of the system within its degenerate ground-state manifold, and the unitary evolution of a system of several MBSs is intimately related to the topology of the braid formed by their worldlines in spacetime \citep{wilczek2021commentary}.  MBSs are thus a type of non-Abelian anyon.  While the theoretical possibility of fractional statistics in spatial dimensions less than three has been known for some time \citep{leinaas&myrheim1977,wilczek1982_fluxTube,laughlin1983}, only recently have experiments begun probing the statistics of Abelian anyons in select fractional quantum Hall states \citep{nakamura2020,bartolomei2020}, while experimental evidence for the existence of non-Abelian anyons remains weak and subject to interpretation.

From an applied perspective, topological superconductors offer a path to intrinsically fault-tolerant quantum information technologies where error correction is implemented at the physical hardware level \citep{kitaev1997_toricCode,kitaev2001_unpairedMajoranas,freedman2002_topoQC,nayak2008review}.  The nonlocal nature of MBSs provides robustness to errors arising due to local noise.  Such local, uncorrelated noise models have been sufficient to describe the error processes in the largest quantum processors to date \cite{google2019_qSupremacy}, suggesting hardware components immune to such errors could provide a more direct route to scalable architectures.  In addition to intrinsically fault-tolerant quantum memories enabled by nonlocality \citep{kitaev2001_unpairedMajoranas}, the non-Abelian statistics of these quasiparticle excitations would enable the implementation of a subset of intrinsically fault-tolerant quantum logic gates needed for universal quantum computation.

In the following we present our perspective of topological superconductivity in superconductor--semiconductor (S--Sm) heterostructures with a focus on planar Josephson junctions in near-surface quantum wells.  In \cref{sec:theory} we present a minimal theory of topological superconductivity and discuss its implementation in different S--Sm hybrid structures.  In \cref{sec:materials} we discuss the primary materials considerations aimed at stabilizing the topological gap and present a number of alternatives to Al--InAs along with their concomitant advantages and disadvantages.  \Cref{sec:topo-or-trivial} presents a major outstanding challenge for the field, namely distinguishing topological and trivial states exhibiting similar experimental signatures.  In \cref{sec:disorder} we focus on disorder as a significant obstacle in S--Sm hybrids and present ideas to better characterize and mitigate its impact.  Finally, in \cref{sec:devices} we discuss a number of promising experimental directions and device integrations.

\section{\label{sec:theory}Quasi-1D Hybrid Systems in 2DEGs}

\begin{figure*}
    \centering
    \includegraphics[width=\linewidth]{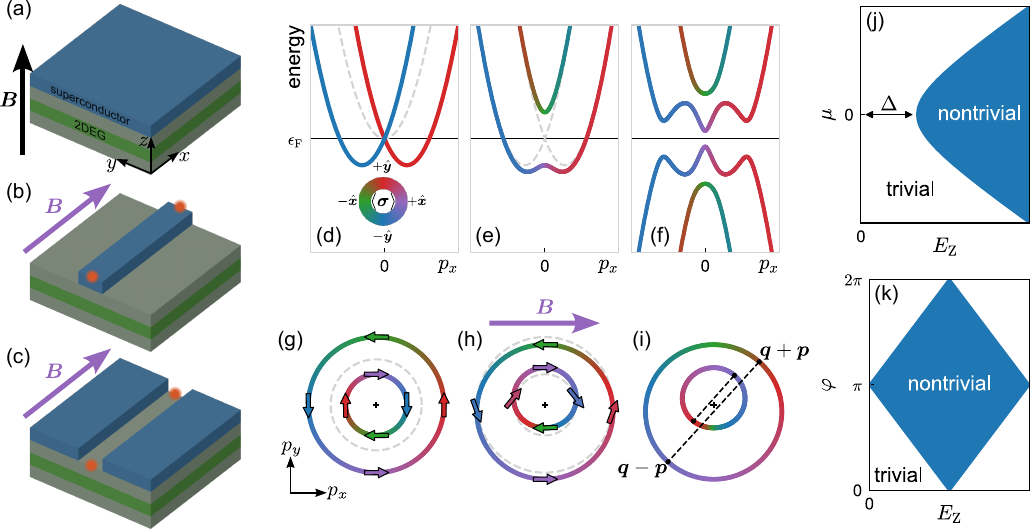}
    \caption[Topo SC in super-semi hybrids]{\textbf{Topological superconductivity in superconductor--semiconductor (S--Sm) hybrids.}  (a,b,c) Geometries for topological superconductivity in S--Sm hybrids.  A superconductor (blue) induces a superconducting gap in a semiconducting two-dimensional electron gas (2DEG, green) by proximity.  The 2DEG is confined to a quantum well with top and bottom barriers (olive green).  A magnetic field $\bm{B}$ is applied.  Coordinate axes corresponding to \eqref{eq:2D-p+ip-hamiltonian} and \eqref{eq:1D-majorana-wire-hamiltonian} are shown in (a).  Majorana bound states are shown schematically as orange circles in (b,c).  (a) Two-dimensional heterostructure in a magnetic field realizing a $p+ip$ topological superconductor described by \eqref{eq:2D-p+ip-hamiltonian}.  (b) A wire and (c) its complement, a planar Josephson junction, realizing (quasi-)one-dimensional p-wave topological superconductors described by \eqref{eq:1D-majorana-wire-hamiltonian}.  The wire is defined by electrostatically gating the underlying 2DEG (not shown).  The color of the magnetic field corresponds to the color wheel in (d).  (d,e,f) Band structure of the 1D Hamiltonian \eqref{eq:1D-majorana-wire-hamiltonian}.  The colors indicate the orientation of the spin expectation value $\expval{\bm{\sigma}}$ in the $xy$-plane according to the color wheel in (d).
    The Fermi energy $\epsilon_\text{F}$ is tuned to the center of the Zeeman gap opened in (e).  (d) spin--orbit coupling splits the parabolic free-particle dispersion (dashed grey) into two spin bands depending on momentum $p_x$.  (e) The Zeeman interaction lifts the $p_x=0$ spin degeneracy in the spin--orbit-split bands (dashed grey).  (f) Superconductivity opens a gap at the Fermi energy.  (g,h,i) Fermi surfaces of the 2D Hamiltonian \eqref{eq:2D-p+ip-hamiltonian} in the $p_xp_y$-plane.  The orientation of the spin expectation value in the $xy$-plane is indicated by both arrows and colors corresponding to the color wheel in (d).  The Fermi energy is tuned above the Zeeman gap for illustrative purposes so that two surfaces are present.  (g) spin--orbit coupling splits the spin-degenerate free-particle surface (dashed grey) into two concentric surfaces with opposite spin textures.  (h) To first order in $E_\text{Z}/\alpha p_\text{F}$, the Zeeman interaction shifts the surfaces in opposite directions along $p_y$ relative to their $E_\text{Z}=0$ positions (dashed grey) and tilts the spin expectation values (colored arrows) in the direction of the field.  (i) Superconducting pairing across the outer Fermi surface is shown schematically for particles with momenta $\bm{q}\pm\bm{p}$ with finite momentum $\bm{q}\parallel-\hat{p}_y$.  (j,k) Topological phase diagrams as a function of Zeeman energy $E_\text{Z}$, with topologically trivial (white) and nontrivial (blue) regimes indicated.  (j) Topological phase diagram \eqref{eq:1D-topological-condition} in $E_\text{Z}$-$\mu$-space for chemical potential $\mu$.  (k) Topological phase diagram in $E_\text{Z}$-$\varphi$-space where $\varphi$ is the phase difference of the superconducting electrodes across the planar junction in (c).}
    \label{fig:hybrids}
\end{figure*}

In this section, we review a general, simplified model of topological superconductivity in two- and one-dimensional S--Sm heterostructures at the level of the Bogoliubov-de Gennes Hamiltonian \citep{lutchyn2010_majoranaNW,oreg2010_majoranaNW,hell2017_planarJJ,pientka2017,alicea2012review,leijnse&flensberg2012intro,chamon} to illustrate the essential physics that will guide our discussion later.  

We begin by confining an otherwise free electron gas to the lowest-energy subband of a quantum well in the $z$-direction, as shown in \cref{fig:hybrids}(a-c).
The confinement is produced by sandwiching a narrow-bandgap semiconductor between two wider-bandgap semiconductors to form the well in the conduction band.
The kinetic energy of the electrons relative to the $z$-confinement subband energy is then
\begin{equation}
    H_\text{K} = \frac{p_x^2 + p_y^2}{2m^*} - \mu
    \label{eq:kinetic-energy}
\end{equation}
with effective mass $m^*$, in-plane momentum $\bm{p}$, and chemical potential $\mu$.

Broken spatial inversion symmetry in the semiconductor, due to e.g.\ the bulk crystal structure or heterostructure, produces a spin--orbit interaction of the general form $\bm{E}\cdot\left(\bm{\sigma}\times\bm{p}\right)$ where $\bm{\sigma}$ is the Pauli vector of spin operators and $\bm{E}$ is the average electric field resulting from the asymmetry.  In an asymmetric well, the dominant contribution to the spin--orbit interaction is commonly due to the structure inversion asymmetry (SIA) of the heterostructure in the $z$-direction, resulting in an electric field $\bm{E} \parallel \hat{\bm{z}}$ and a spin--orbit interaction of the Bychkov-Rashba form \citep{bychkov&rashba1984_2deg,fabian2007review}
\begin{equation}
    H_\text{R} = \alpha\left(\sigma_xp_y - \sigma_yp_x\right)
    \label{eq:spin--orbit-interaction}
\end{equation}
where the coupling strength $\alpha$ depends on the electric field in the valence band of the quantum well and can be further modulated by electrostatic gating; see \citet{fabian2007review} for details.  For simplicity, we neglect additional terms such as Dresselhaus spin--orbit coupling which may be present due to bulk inversion asymmetry (BIA).  The effect of \cref{eq:spin--orbit-interaction} on the free-electron dispersion is to lift the spin degeneracy at all points in the Brillouin zone except for $\bm{p}=0$, as shown in \cref{fig:hybrids}(d).

Lifting the remaining degeneracy at $\bm{p}=0$ to induce (partially) spin-polarized bands, as required for p-wave pairing in topological superconductors, requires breaking time-reversal symmetry.  This may be done e.g.\ by an exchange interaction with a ferromagnetic insulator or by applying a magnetic field $\bm{B}$ with the resulting Zeeman interaction
\begin{equation}
    H_\text{Z} = -\frac{1}{2}g^*\mu_\text{B}\bm{B}\cdot\bm{\sigma},
    \label{eq:zeeman-interaction}
\end{equation}
where $g^*$ is the effective $g$-factor and $\mu_\text{B}$ the Bohr magneton.  The direction of the magnetic field should be chosen perpendicular to the spin--orbit field and therefore depends on both the relative strength of Rashba and Dresselhaus spin--orbit interactions and the crystallographic orientation of the system \citep{scharf2019,pakizer2021,pekerten2022}.  The Zeeman interaction opens a gap at $\bm{p}=0$, as shown in \cref{fig:hybrids}(e) and mixes the spin eigenstates such that superconducting correlations that favor pairing of anti-parallel spins of opposite momenta can produce Cooper pairs of electrons with finite parallel spin projection.

Finally, we introduce an s-wave or BCS-like pairing term,
\begin{equation}
    H_\text{S} = \Delta \tau_+ + \Delta^*\tau_-,
    \label{eq:s-wave-superconductivity}
\end{equation}
where $\Delta$ is the complex superconducting order parameter induced in the underlying 2DEG by proximity coupling to a thin-film superconductor, and $\tau_\pm = \frac{1}{2}(\tau_x \pm i\tau_y)$ where the $\tau_j$ are the Pauli operators in electron-hole space.  The pairing interaction opens a gap at the Fermi level, and the introduction of the Nambu basis doubles the spectrum, as shown in \cref{fig:hybrids}(f).

Altogether, the ingredients of \cref{eq:kinetic-energy,eq:spin--orbit-interaction,eq:zeeman-interaction,eq:s-wave-superconductivity}, with the magnetic field in the $z$-direction,
realize a two-dimensional $p_x\pm ip_y$ superconductor.
In the conventional Nambu basis $\Psi^\dag = (\psi_\uparrow^\dag, \psi_\downarrow^\dag, \psi_\downarrow, -\psi_\uparrow )$,
\begin{equation}
    H_\text{2D} =
    \left[\xi_p + \alpha\left(\sigma_xp_y - \sigma_yp_x\right)\right]\tau_z - E_\text{Z}\sigma_z + \Delta\tau_+ + \Delta^*\tau_-
    \label{eq:2D-p+ip-hamiltonian}
\end{equation}
with Zeeman energy $E_\text{Z}=\frac{1}{2}g^*\mu_\text{B}B$ and kinetic energy $\xi_p=\bm{p}^2/2m^* - \mu$.  At this point, the topology of the system is characterized by a Chern number that counts the number of chiral Majorana edge modes.

If we further confine the system to a bound state in the $y$-direction (see \cref{fig:hybrids}(b,c)), $\expval{p_y} = 0$ and we arrive at the Hamiltonian for a Majorana wire \citep{lutchyn2010_majoranaNW,oreg2010_majoranaNW},
\begin{equation}
    H_\text{1D} =
    \left(\xi_p - \alpha \sigma_yp_x\right)\tau_z - E_\text{Z}\sigma_x + \Delta\tau_+ + \Delta^*\tau_-,
    \label{eq:1D-majorana-wire-hamiltonian}
\end{equation}
where we have chosen to align the magnetic field in the experimentally more practical $x$-direction because the critical field of the parent superconductor is largest in the plane of the thin film.  In a gate-defined wire (\cref{fig:hybrids}(b)), the confinement is achieved by depleting the 2DEG outside the proximitized region covered by the superconducting strip, while in a planar junction (\cref{fig:hybrids}(c)) coherent multiple Andreev reflections between the two superconductor-normal interfaces constructively interfere to form Andreev bound states (ABSs).  Diagonalizing the $4\times 4$ Hamiltonian \eqref{eq:1D-majorana-wire-hamiltonian} yields the particle-hole symmetric energy dispersion, and considering the low-energy eigenvalue shows that the spectrum is gapless at $p_x=0$ when $E_z^2 = \abs{\Delta}^2 + \mu^2$.  This defines a phase boundary between topologically trivial and nontrivial phases, as shown in \cref{fig:hybrids}(j).  The system is topologically nontrivial for
\begin{equation}
    E_z^2 > \abs{\Delta}^2 + \mu^2.
    \label{eq:1D-topological-condition}
\end{equation}

The Hamiltonian \eqref{eq:1D-majorana-wire-hamiltonian} is characterized by a $\mathbb{Z}_2$ topological index that identifies the ground state degeneracy \cite{altland&zirnbauer1997,ryu2010}.  In the topologically trivial phase, there is a unique ground state of even fermion parity corresponding to a superconducting condensate with all electrons paired.  In the topologically nontrivial phase, the ground state is doubly degenerate\footnote{The energy splitting is not exactly zero, but rather exponentially small with respect to the system size relative to the coherence length.} due to the existence of a pair of MBSs, localized at the two ends of the system, that can accommodate a single electron at zero energy cost.  The $\mathbb{Z}_2$ index can be illustrated in a simplified manner by cyclically permuting the coordinate axes and considering the gap near $p=0$.  For small momenta, we can neglect the kinetic energy term quadratic in $p$ (and taking $\mu=0$) to obtain
\begin{equation}
    H_\text{1D} \approx \alpha p_z \sigma_x \tau_z - E_\text{Z}\sigma_z + \Delta\tau_+ + \Delta^* \tau_-
    \label{eq:1D-Hamiltonian-near-p=0}
\end{equation}
and
\begin{equation}
    H_\text{1D}^2 \approx (\alpha p_z)^2 + E_\text{Z}^2 + \abs{\Delta}^2 - 2E_\text{Z}\sigma_z\left(\Delta\tau_+ + \Delta^*\tau_-\right).
    \label{eq:1D-Hamiltonian-near-p=0-squared}
\end{equation}
According to \eqref{eq:1D-Hamiltonian-near-p=0-squared}, the low-energy subspace is spanned by the two eigenstates of $\sigma_z\left(\Delta\tau_+ + \Delta^*\tau_-\right)$ with eigenvalue $+\abs{\Delta}$.
Projecting \eqref{eq:1D-Hamiltonian-near-p=0} onto this subspace yields an effective low-energy Dirac Hamiltonian
\begin{equation}
    H_\text{1D} \approx \left(\abs{\Delta} - E_\text{Z}\right)\tilde\tau_z + \alpha p_z\tilde\tau_x
\end{equation}
with Dirac mass $m_\text{D}=\abs{\Delta}-E_\text{Z}$.  We see that the Zeeman interaction and superconductivity compete to gap out the spectrum at $p=0$, and a topological phase transition occurs at the band inversion where the Dirac mass changes sign.  The system topology is then indicated by the sign of the Dirac mass, with $\sign m_\text{D}=+(-)1$, corresponding to trivial(nontrivial) topology.

In the planar junction geometry (see \cref{fig:hybrids}(c)), the separation of the superconducting electrodes enables a phase difference $\varphi$ to occur between them.  This phase is used as an additional experimental tuning parameter e.g.\ by current biasing the junction or by flux biasing a superconducting loop in which the junction is embedded.  The effect of the superconducting phase difference on the topological phase diagram is schematically shown in \cref{fig:hybrids}(k).

The above description does not include a number of important experimental details: 1)
The proximity coupling and confinement lead to a renormalization of the material parameters $\alpha$ and $g^*$ away from their bulk semiconductor values. 2) We assumed a perfectly clean system free of any disorder. 3) Magnetism suppresses and eventually closes the parent superconducting gap. 4) The details of the proximity effect determine the magnitude and hardness of the superconducting gap induced in the semiconducting 2DEG; the model does not include interface information.  5) We assumed a purely one-dimensional system occupying the lowest confinement subband. 6) We have not included the orbital effect of the magnetic field. Contributions from the above experimental details impact the possibility of realizing topological superconductivity, the properties of the resulting Majorana wavefunctions, and their experimental signatures.  In the following, we discuss these experimental details with a focus on epitaxial S--Sm platforms.

\section{\label{sec:materials} Epitaxial Super/Semi Materials}

\begin{figure*}
    \centering
    \includegraphics[width=\linewidth]{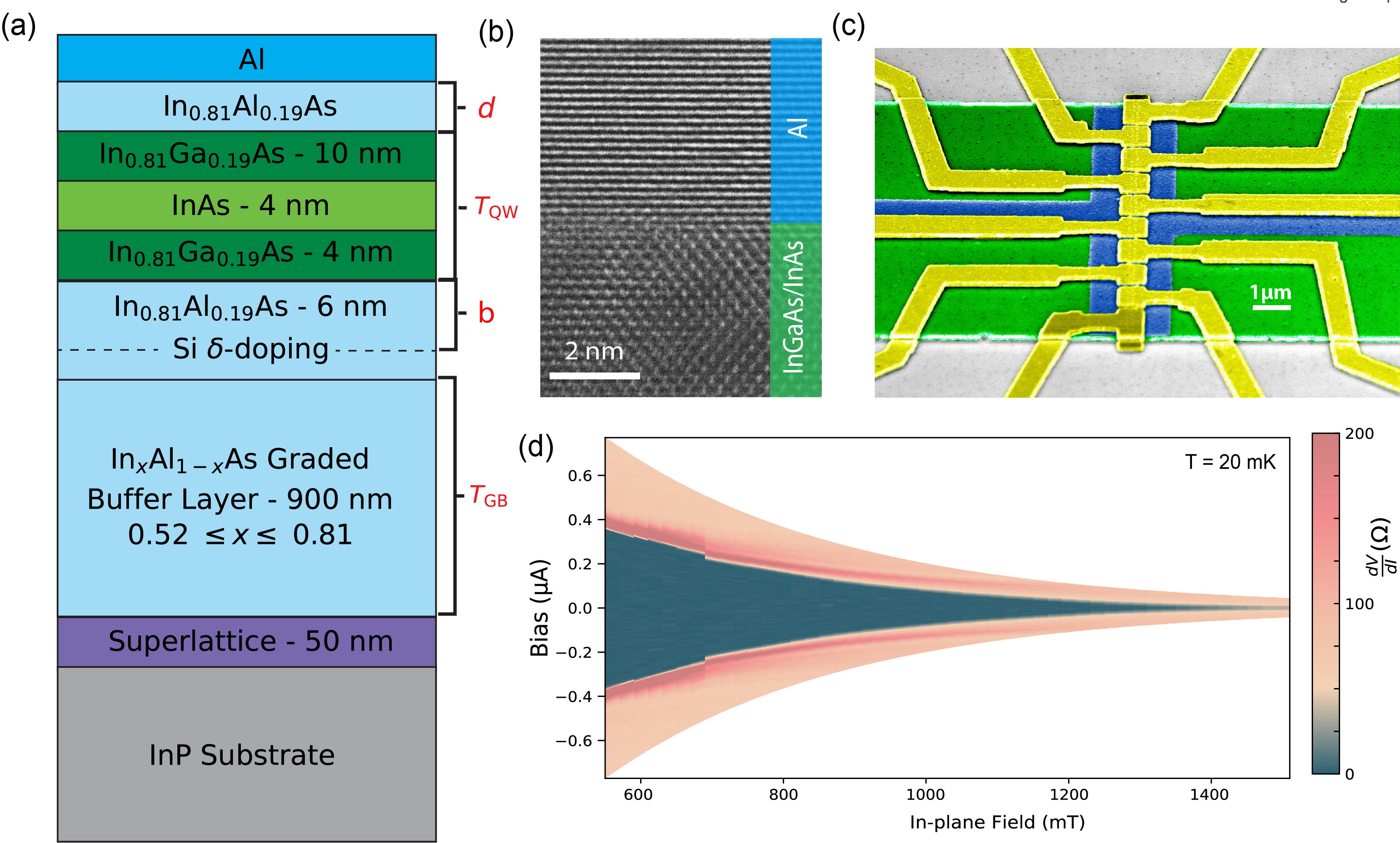}\\
    \caption[clean super-semi interface]{\textbf{Engineering clean superconductor--semiconductor (S--Sm) interfaces.} (a) An InGaAs/InAs/InGaAs quantum well is grown atop an InP substrate by lattice matching via a step-graded \ce{In_xAl_{1-x}As} buffer layer.  A thin \ce{In_{0.81}Al_{0.19}As} layer serves as a diffusion barrier between the Al and the quantum well \citep{sarney2018,sarney2020}.  (b) Transmission electron micrograph of an atomically clean S--Sm interface enabling a hard induced gap.  (c) Gated, planar Josephson junction fabricated on the heterostructure in (a). Multiple independently tunable gates are designed to spatially modulate the topology of the junction via the chemical potential in \eqref{eq:1D-topological-condition} \citep{elfeky2021_minigate}. (d) Differential resistance vs.\ in-plane field showing the large in-plane critical field of the junction due to the thin superconducting film.}
    \label{fig:disorder:S-Sm}
\end{figure*}

\subsection{Wires and wells}
For a decade, the realization of S--Sm hybrid structures for topological superconductivity in one dimension relied on self-assembled nanowires and accelerated with the advent of epitaxial interfaces between Al and InAs nanowires \cite{krogstrup2015}.  Recently, the experimental focus has shifted from nanowires towards quasi-one-dimensional systems patterned in two-dimensional electron gases (2DEGs) confined to quantum wells \cite{shabani2016}, including gate-defined wires \cite{nichele2017_scalingZBCP,suominen2017_zbp,ofarrell2018,lee2019_transport,poschl2022_2,microsoft2023,microsoft2024}
and planar Josephson junctions \cite{shabani2016,mayer2019_proximitizedInAs,ren2019,lee2019_transport,pankratova2020,mayer2020_proximitizedInAsSb,mayer2020_anomalousPhase,dartiailh2021_piJump,dartiailh2021_shapiro}. 
In these systems, MBSs are still predicted to appear at the two ends of a quasi-1D channel (see \cref{fig:hybrids}), in contrast to the MBSs bound to the normal-state vortex cores in 2D chiral $p_x \pm ip_y$ superconductors \citep{read&green2000}.

The most developed epitaxial S--Sm system is based on the well-established epitaxial relationship between Al and InAs quantum wells \cite{shabani2016}. \Cref{fig:disorder:S-Sm}(a) shows a typical InAs quantum well material stack on an InP (100) substrate. The stack starts with a step graded buffer layer of In$_{x}$Al$_{1-x}$As that is grown at low temperature, typically $T_\text{GB} \sim$\SI{360}{\celsius}, to minimize dislocations forming due to the lattice mismatch between the active region and the InP substrate.  The quantum well is typically grown at a higher temperature $T_\text{QW} \sim \SI{420}{\celsius}$ and consists of a \SI{4}{nm} layer of InAs grown on a \SI{4}{nm} layer of In$_{0.81}$Ga$_{0.19}$As.  A top layer, typically \SI{10}{nm} of In$_{0.81}$Ga$_{0.19}$As, is grown on the strained InAs quantum well as shown. An extra layer of In$_{0.81}$Al$_{0.19}$As, $d \approx \SI{1}{nm}$ can be grown on top to reduce the coupling of the superconductor and the 2DEG if desired. The structure is usually delta doped with Si $b = \SI{6}{nm}$ below the quantum well. After the quantum well is grown, the substrate is cooled to promote the growth of epitaxial Al (111) \cite{shabani2016}.  \Cref{fig:disorder:S-Sm}(b) shows a high-resolution transmission electron microscope (TEM) image of this interface between Al and In$_{0.81}$Ga$_{0.19}$As, with atomic planes of both crystals clearly visible.  A 2-inch wafer with epitaxial Al is processed by selectively removing the Al using Transene Type-D Al etchant to define device features. \Cref{fig:disorder:S-Sm}(c) shows a false colored scanning electron micrograph of a Josephson junction formed by removing \SI{100}{nm} by \SI{8}{\micro m} of Al. Using the layer by layer growth of epitaxial Al, it is possible to stop the growth at ultra thin thicknesses (considering the Al oxidation of \SIrange{1}{2}{nm} after exposure to ambient conditions). It is routine to target an Al thickness of \SI{10}{nm} or less for enhanced in-plane critical magnetic field. \Cref{fig:disorder:S-Sm}(d) shows supercurrent as a function of applied parallel magnetic field exhibiting \SI{1.4}{T} critical field.
%


These proximitized quantum wells offer a number of improvements as well as challenges when compared to nanowires.  For example, these systems are more amenable to top-down design and nanofabrication using well-established wafer-scale lithographic techniques, enabling more flexible, scalable, and creative device geometries (see \cref{sec:devices}).
However, the material processing involved, particularly wet etching and dielectric growth, introduces additional disorder to the system.
The necessity of a substrate material for growth and the resulting misfit and dislocations in quantum wells also creates unwanted interface and surface roughness.

In addition, the thickness of the top barrier of the quantum well provides a natural and important experimental tuning knob which is absent in the nanowire system.
The top barrier thickness controls the strength of the proximity coupling between the semiconducting 2DEG and the superconductor.
Several theoretical studies identify numerous adverse effects of too-strong S--Sm coupling, including a larger topological critical field and a renormalization of the semiconductor g-factor and spin--orbit coupling \citep{stanescu2011,cole2015,stanescu&dassarma2017,reeg2018_metallize,awoga2022_weakCoupling}.
On the other hand, stronger coupling typically implies a shorter MBS localization length.
The top barrier also has implications for topological superconductivity in the presence of disorder (see \cref{sec:disorder}) and helps to mitigate surface scattering, yielding higher-mobility electron gases \cite{shabani2014_gating,yi2015,suto2022,wickramasinghe2018} than in nanowires.
Thus, the top barrier helps to manage complex trade-offs regarding S--Sm proximity coupling, disorder and mobility.
Recent work has implemented an analogous tunnel barrier in InSb nanowires \citep{badawy2023}.


\subsection{\label{sec:materials:gap}Superconducting Gap: Enhancement vs.\ Preservation}
Only a few superconductors have been explored for topological Josephson junctions, most notably epitaxial Al due to the advantages of its self-limiting native oxide and a hard induced gap in proximate semiconductors \cite{krogstrup2015,chang2015,shabani2016}. Al is also widely known to exhibit charge parity stability demonstrated by the 2e charging in small islands \cite{albrecht2016,shen2018}. Among the disadvantages of Al are its relatively small superconducting gap, critical temperature and critical field \cite{chang2015}. This restricts the design of future topological qubits requiring a precise balance of several energy scales \cite{pan2020_goodBadUglyZBCPs}. While the critical field of epitaxial Al is enhanced by reducing the Al thickness, the large surface scatterings and unignorable nonuniformity in the thin Al film could lead to reduced mean free path and coherence length \cite{mazur2022_AlPt,levajac2023_jjLength}. Due to these limitations with Al, there is strong interest in broadening the scope of superconductor/semiconductor hybrids. Several works explored the growth of Sn, Pb, and Nb compounds and alloys on InAs and InSb nanowires and showed promising results exhibiting a large, hard induced gap resilient to high magnetic fields and elevated temperatures \cite{pendharkar2021,kanne2021,gul2017,drachmann2017}. These new hybrids also manifested the coveted two-electron charging effect in small islands \cite{pendharkar2021,kanne2021}, indicating a charge parity stability comparable to Al.  A hard gap in germanium induced by superconducting PtSiGe formed by annealing has also recently been demonstrated \citep{tosato2023} as well as in InAsSb with epitaxial Al \citep{sestoft2018}.

In \cref{fig:material-choices} we plot the topological critical points for various combinations of semiconductors and superconductors.
As the size of the topological gap deep within the topological regime is ultimately limited by the parent gap of the proximitizing superconductor, one might think at first that a large-gap superconductor like Nb and its compounds \citep{drachmann2017,aggarwal2021} would be preferable to a superconductor like Al, the most common superconductor of choice for engineering topological superconductivity.
However, an increase in the parent gap without a concomitant increase in g-factor results in a larger topological transition magnetic field.
At these large fields, the electron spins become increasingly polarized, which actually decreases the effectiveness of the s-wave pair potential, which can only bind electrons of opposite spin.
In the Kitaev limit where the Zeeman energy dominates over the spin--orbit, $\alpha$, the effective p-wave potential goes as $\sim\Delta\alpha/E_Z$ \citep{chamon}.
Hence any increase in $\Delta$ is offset by a larger required Zeeman energy.
On the other hand, the topological gap could be increased by using materials with stronger spin--orbit coupling.

The important and relevant $\Delta$ for topological superconductivity is really the proximity-induced gap in the semiconductor, which can approach the value of the parent gap given a sufficiently clean and transparent interface.
However, large-gap superconductors have smaller coherence lengths which require shorter weak links to remain in the short junction limit $L \ll \xi$, where $L$ is the distance between the superconducting contacts and $\xi$ is the (renormalized) superconducting coherence length induced in the underlying semiconductor.  In long junctions ($L \gg \xi$), subgap states exist in the weak link for all phase biases and limit the ultimate size of the topological gap.
In addition, larger superconducting gaps necessitate smaller device dimensions $L$ in a phase-winding geometry (\cref{sec:phase}) by a rule of thumb $L\Delta \sim \alpha$ \citep{lesser2021_3phase,lesser2021_phaseInduced}.
Finally, achieving transparent interfaces is challenging with certain S--Sm combinations with incompatible crystal structures or incommensurate lattice constants; see \cref{sec:disorder:interface}.  In \cref{tab:superconductor-material-parameters} we tabulate crystallographic and superconducting parameters for various superconductors.  We note that the induced gap can be on the order of the parent gap, while the topological gap is typically reduced by an order of magnitude.  Therefore, the parent gap should be chosen at the very least an order of magnitude larger than the $k_\text{B}T \approx \SI{1}{\micro eV}$ floor of a typical \SI{10}{mK} dilution refrigerator, and in practice much larger. 
The most important criterion is that the superconductor induce a sizeable gap in the semiconductor that is free of subgap states, i.e.\ a hard gap.  This requires a clean S--Sm interface such as Al--InAs; see \cref{sec:disorder:interface}. As seen in \cref{tab:superconductor-material-parameters} there are several superconductors with larger superconducting gap compared to Al. For example A15 family provides an enhancement of 18 times the superconducting gap. What remains to be explored is whether it is possible to achieve clean interfaces with semiconductors of strong spin--orbit coupling for enhanced topological gap. 

\begin{table*}[tbp]
    \centering
\begin{tabular}{llSSSSSS}
\toprule
{material} & {crystal} & {a (\si{\angstrom})} & {b (\si{\angstrom})} & {c (\si{\angstrom})} & {$T_c$ (K)} & {$\Delta$ (meV)} & {$\xi$ (nm)} \\
\midrule
\ce{Nb_3Ge} & A15 & 5.156 &  &  & 23.2 & 3.527 & 59 \\
\ce{Nb_3Ga} & A15 & 5.168 &  &  & 20.7 & 3.147 & 67 \\
\ce{Nb_3Al} & A15 & 5.180 &  &  & 19.1 & 2.903 & 72 \\
NbN & diamond & 4.394 &  &  & 16.0 & 2.432 & 86 \\
\ce{Nb_3Au} & A15 & 5.2027 &  &  & 11.5 & 1.748 & 120 \\
\ce{Nb_3Pt} & A15 & 5.1547 &  &  & 10.9 & 1.657 & 126 \\
NbTi &  &  &  &  & 9.6 & 1.459 & 144 \\
Nb & bcc & 3.304 &  &  & 9.25 & 1.406 & 149 \\
Pb & fcc & 4.9508 &  &  & 7.22 & 1.098 & 191 \\
Ga:Ge \citep{strohbeen2023_hyperdopedGe} &  &  &  &  & 5.27 & 0.801 & 262 \\
Ta & bcc & 3.3019 &  &  & 4.47 & 0.679 & 308 \\
$\beta$-Sn & tetrg & 5.8316 &  & 3.1815 & 3.722 & 0.566 & 370 \\
In & tetrg & 3.2530 &  & 4.9455 & 3.408 & 0.518 & 404 \\
\ce{TaGe} \citep{strohbeen2024_TaGe} &  &  &  &  & 1.9 & 0.289 & 725 \\
Al & fcc & 4.0495 &  &  & 1.175 & 0.179 & 1173 \\
Ga & orthr & 4.523 & 7.661 & 4.524 & 1.083 & 0.165 & 1273 \\
PtSi & orthr & 5.595 & 3.603 & 5.932 & 0.88 & 0.134 & 1566 \\
Zn & hcp & 2.6646 &  & 4.49455 & 0.85 & 0.129 & 1622 \\
PtSiGe \citep{tosato2023} &  &  &  &  & 0.5 & 0.076 & 2757 \\
PtGe & orthr & 5.733 & 3.701 & 6.088 & 0.4 & 0.061 & 3446 \\
\bottomrule
\end{tabular}
\caption{Material parameters for various superconductors.  Lattice parameters $a$, $b$, $c$, and critical temperature $T_\text{c}$ were compiled from Refs.~\onlinecite{poole,arblaster}.  The superconducting gap is estimated from the BCS relation $\Delta = 1.764 k_\text{B} T_\text{c}$.  The coherence length of the induced gap is estimated from the BCS relation $\xi = \hbar v_\text{F} / \pi\Delta$ using $v_\text{F} = \SI{e6}{m/s}$ typical for semiconductors and corresponding to a carrier density $\sim\SI{7e11}{cm^{-2}}$ in InAs.}
    \label{tab:superconductor-material-parameters}
\end{table*}

As the system's topological properties are relatively less sensitive to the magnitude of superconducting gap than other factors like the disorder in the semiconductor, super-semi interface, and semi-oxide interface, increasing the parent gap is not necessarily the highest priority \citep{sticlet2017}.
A more immediate strategy could be focused to preserve the existing parent gap at higher field. This might be done by doping the superconductor at its surface with a heavy element like platinum to provide spin-relaxation channels via spin--orbit scattering in the superconductor with minimal impact on the semiconductor \citep{tedrow&meservey1979,tedrow&meservey1982,mazur2022_AlPt,vanloo2023}.
Another promising direction is to reduce the critical Zeeman field (or eliminate it entirely) by leveraging alternative sources of T-symmetry breaking, such as supercurrent \citep{melo2019} and phase bias \citep{fu&kane2008,vanHeck2014,hell2017_planarJJ,pientka2017,ren2019,fornieri2019,lesser2021_3phase,lesser2021_phaseInduced,lesser2022}; see \cref{sec:phase}.

\subsection{\label{sec:materials:soc}Spin--Orbit Coupling in Proximity Devices}

Spin--orbit coupling is understood to be the most important parameter to topological superconductivity in hybrid systems, setting upper limits on the topological gap and all other energy scales of the system in the topological regime \citep{sau2012}.
However, strengthening the spin--orbit coupling can often come at a cost.
For example, certain confinement effects renormalize the magnitudes of spin--orbit coupling and g-factor (\cref{sec:g-factor}) in opposite directions.
Narrowing the quantum well enhances the interfacial contribution to the Rashba spin--orbit coupling and increases the linearized Dresselhaus coefficient \citep{mayer2020_proximitizedInAsSb,walser2012,fabian2007review}, but band structural effects associated with confinement such as an increased band gap and nonparabolic dispersion effects can suppress the g-factor \citep{smithiii&fang1987,mayer2020_proximitizedInAsSb,yuan2020}.
On the other hand, at low densities electron--electron interactions become more relevant and have the opposite effect, enhancing g-factor while suppressing spin--orbit coupling \citep{maryenko2021}.
A trade-off between g-factor and spin--orbit coupling must therefore be weighed when choosing the quantum well width, with spin--orbit coupling being maximized to the extent that g-factor can be sacrificed (see \cref{fig:material-choices}).
All other strategies should likewise be employed to increase the spin--orbit coupling in the system, from material choice to gate-tunability \citep{wickramasinghe2018,farzaneh2024} and device design (\cref{sec:periodic}).

Given the various factors, both material and structural, impacting the effective spin--orbit coupling in a heterostructure, due diligence must be done in characterizing each wafer before any detailed understanding of individual devices fabricated from that material can be obtained.
Weak antilocalization measurements in a Hall bar geometry are a straightforward way to estimate the spin--orbit coupling strength in a given semiconducting material, but extracting quantitatively accurate results that can be compared across disparate material systems is challenging.
The weak antilocalization peak in the magnetoconductance is caused by the Berry phase associated with the spin--orbit field, which yields destructive interference of closed electron trajectories and a suppression of backscattering.
This peak is therefore a signal of the spin--orbit coupling strength that can be studied as a function of the controllable system parameters.
In particular, the Rashba parameter, a material-dependent parameter usually dominated by the structural inversion asymmetry of the quantum well, is modulated by tilting the bands with an electric field applied by a gate \citep{wickramasinghe2018,farzaneh2024}.
Moreover, weak antilocalization studies in the presence of an in-plane magnetic field reveal an anisotropy that arises due to non-negligible Dresselhaus spin--orbit coupling, with implications for the optimal in-plane field angle to induce topological superconductivity in planar junctions \citep{scharf2019,pakizer2021,pekerten2022,farzaneh2024}.

The relevant material parameters are further renormalized in devices due to semiconductor-superconductor hybridization and additional confinement effects.
In-situ characterization of the material parameters within the device itself is therefore imperative.
Measurement of the junction's current-phase relation may reveal an anomalous phase shift due to the combined effects of the Zeeman and spin--orbit fields \citep{mayer2020_anomalousPhase}:
\begin{equation}
    \varphi_0 \propto g\mu_\text{B} (\alpha^2 - \beta^2)(B_x \alpha + B_y \beta),
\end{equation}
where $\beta$ is the Dresselhaus spin--orbit coupling strength.
Again, gate-control of $\alpha$ proves critical in separating the contributions of the Zeeman effect and spin--orbit coupling.
The microscopic superconducting diode effect \citep{lotfizadeh2024,baumgartner2021_diode,baumgartner2022,costa2023_diodeSignReversal,bauriedl2022,reinhardt2024,yokoyama2013,yokoyama2014,ando2020,legg2022,turini2022,mazur2022_diode} likewise reveals the interplay between the Zeeman and spin--orbit fields and can be used to characterize the spin--orbit coupling strength acting in the device, though care should be taken to account for alternative coexisting mechanisms \citep{lotfizadeh2024}.
Here, s-wave superconductivity binds Cooper pairs with finite momentum across the Rashba- and Zeeman-split bands \citep{lotfizadeh2024,pal2022_diode}, as shown in \cref{fig:hybrids}(i).

\begin{table*}[tbp]
    \centering
    \begin{threeparttable}
\begin{tabular}{lSS[table-format=+1.5]S[table-format=3.3]SSS[table-format=+2.2]S}
\toprule
{material} & {a (\si{\angstrom})} & {$m^*/m_e$} & {$\alpha$ (\si{meV.\angstrom})} & {$\beta$ (\si{meV.\angstrom})} & {$E_\text{so}$ (\si{\micro eV})} & {$g^*$} & {$E_\text{Z}/B$ (\si{\micro eV/T})} \\
\midrule
graphene &  & 0 &  &  &  & 1.95\tnote{d} & 56 \\
Si & 5.4310 & 0.188 &  &  &  & 2.0 & 57 \\
Ge & 5.6579 & 0.038\tnote{a} &  &  &  & -3.2\tnote{e} & -93 \\
AlAs & 5.66139 & 1.1\tnote{b} & 0.43 & 7.91 & 20.1 & 1.9 & 56 \\
GaAs & 5.65330 & 0.067 & 4.72 & 16.75 & 8.1 & 0.0 & -1 \\
GaSb & 6.09593 & 0.039 & 35.52 & 122.35 & 255.1 & -8.8 & -253 \\
InP & 5.8690 & 0.07927 & 1.57 & -7.09 & 1.6 & 1.3 & 38 \\
InAs & 6.0583 & 0.024 & 112.49 & 33.33 & 133.9 & -14.9 & -431 \\
InSb & 6.47937 & 0.013 & 534.21 & 324.60 & 2516.6 & -57.5 & -1664 \\
InAsSb & 6.26883 & 0.011 & 325\tnote{c} &  & 305.0 & -69.2 & -2001 \\
ZnSe & 5.6692 & 0.137 & 1.057 &  & 0.0 & 1.3 & 38 \\
CdTe & 6.481 & 0.090 & 6.930 &  & 1.1 & -1.2 & -34 \\
HgTe & 6.4603 & -0.028 &  &  &  & -27.2\tnote{f} & -786 \\
\bottomrule
\end{tabular}
\begin{tablenotes}
    \item[a] The L valley is slightly lower than the $\Gamma$ valley \citep{clavel2022}, with $m^*_\text{t} = 0.081m_\text{e}$ and $m^*_\text{l} = 1.61m_\text{e}$ \citep{adachi_semiconductors}.
    \item[b] $m^*=0.124m_\text{e}$ at $\Gamma$ point.  Electrons occupy the X valleys \citep{shayegan2006} with $m^*_\text{t} = 0.19m_\text{e}$ and $m^*_\text{l} = 1.1m_\text{e}$ \citep{adachi_semiconductors}.
    \item[c] Refs.~\onlinecite{mayer2020_proximitizedInAsSb,moehle2021}.
    \item[d] \citep{lyon2017}
    \item[e] At the $\Gamma$ point.  $g^*\lesssim 4$ in the L valley \citep{roth1959,roth&lax1959,watzinger2016}.
    \item[f] Up to -90 observed \citep{jiang2023_gFactor}.
\end{tablenotes}    \end{threeparttable}
    \caption{Material parameters for various zinc-blende semiconductors.  The lattice parameters $a$ and electron effective masses $m^*$ at the $\Gamma$ point were taken from Refs.~\onlinecite{adachi_semiconductors,adachi_alloys}.  Rashba parameters $\alpha$ and linearized Dresselhaus parameters $\beta$ are results of (extended) Kane model calculations from Refs.~\onlinecite{fabian2007review,winkler}.  The spin--orbit energy $E_\text{so}=2m^*\left((\abs{\alpha}+\abs{\beta})/\hbar\right)^2$ is estimated for a Fermi level within the $p=0$ Zeeman gap (see \cref{fig:hybrids}).  The effective g-factor $g^*$ is estimated from the Roth formula \citep{mayer2020_proximitizedInAsSb,roth1959} using band parameters from Refs.~\onlinecite{adachi_semiconductors,adachi_alloys}, from which the Zeeman energy per unit field, $E_\text{Z}/B = g^*\mu_\text{B}/2$ is calculated.}
    \label{tab:semiconductor-material-parameters}
\end{table*}


\subsection{Importance of \label{sec:g-factor}$g$-factor}

\begin{figure}
    \centering
    \includegraphics{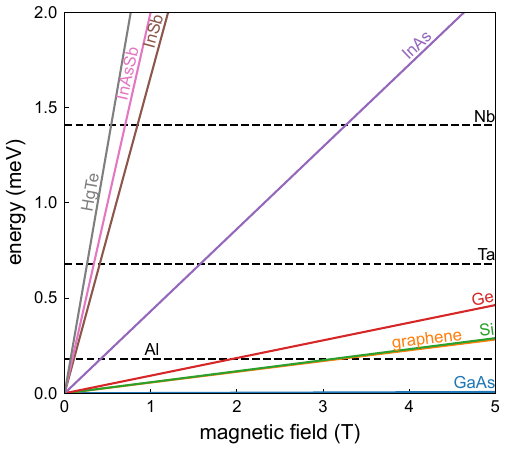}
    \caption[topological condition]{\textbf{Impact of material parameters on topological critical field.}  Zeeman energy $\abs{g^*}\mu_\text{B} B / 2$ for various semiconductors (colored solid lines) and superconducting gap $\Delta$ for various superconductors (black dashed lines).  A topological phase transition occurs at the intersection of each superconductor--semiconductor pair according to \eqref{eq:1D-topological-condition} (with $\mu=0$).}
    \label{fig:material-choices}
\end{figure}

The effective $g$-factor of the confined, proximitized electron gas is an important material parameter when an external magnetic field is used as the source of time-reversal symmetry breaking.\footnote{
Broken T-symmetry is required to obtain spatially isolated, unpaired Majorana quasiparticles; otherwise, Kramers degeneracy theorem implies the existence of a second, time-reversed pair of Majorana modes that overlaps the first \citep{leijnse&flensberg2012intro}.
}
While the Zeeman field is required to open a topological gap according to \eqref{eq:1D-topological-condition}, it also suppresses the s-wave gap of the parent superconductor.
A large g-factor is therefore desirable to attain larger Zeeman energies at smaller magnetic fields; see \cref{fig:material-choices}.
While many materials with large g-factors do exist (see \cref{tab:semiconductor-material-parameters}), the g-factor is generally renormalized to smaller values when the charge carriers are confined \citep{mayer2020_proximitizedInAsSb}.  
Small g-factors are a challenge for Majorana platforms based on group IV semiconductors such as germanium which otherwise offer relatively low disorder and a hard superconducting gap \citep{tosato2023,luethi2023,laubscher2024_MZMGeNW}.

\section{\label{sec:topo-or-trivial}Distinguishing topological and trivial states}
\subsection{Trivial States in Hybrid Systems}

While S--Sm hybrid systems are predicted to host MBSs, they are also naturally the nest of many kinds of non-Majorana states.
Among these topologically trivial states, Andreev bound states (ABSs) are the focus of many theoretical and experimental works as they are precursors which evolve into MBSs at the topological phase transition.
At a normal-metal--superconductor interface, an incoming electron in the normal metal can be retroreflected as a hole to form a Cooper pair in the superconductor in a process called Andreev reflection.
In a superconductor--normal-metal--superconductor junction, multiple Andreev reflections between the two interfaces interfere to form ABSs which mediate the flow of supercurrent across the junction.
More generally, however, the formation of ABSs requires only some degree of spatial variation of the superconducting pair potential, which could happen in systems with magnetic domains or impurities (Yu-Shiba-Rusinov states).
ABSs have been carefully studied in hybrid nanowire systems as well as planar Josephson junctions using tunneling spectroscopy \cite{deacon2010,lee2014_spinResolved,su2018,juenger2019,nichele2020} and microwave spectroscopy \cite{vanWoerkom2017,hays2018,chidambaram2022,hinderling2023_absSpec}
and are known to mimic the signatures of their topological MBS counterparts. Especially in a system with strong SO coupling and high electron density, the suppressed g-factor \cite{vanHeck2017} could cause ABSs to have extended zero-energy pinning \cite{lee2014_spinResolved} forming zero-bias conductance peaks (ZBCPs) that are difficult to distinguish from those of an MBS.

When taking disorder into account, the zoo of trivial ABSs is further expanded. It has been shown that the combination of multiple subbands with disorder can also give rise to ABSs in class D Hamiltonians, which leads to the emergence of so-called `class-D peaks'. As the results of level repulsion from higher energy states under the influence of a Zeeman field, class-D peaks can form ZBCPs but with limited zero energy pinning \cite{prada2017}. However, it is still possible to have prolonged class-D peaks through fine-tuning of parameters and data selection.

Recently, ABSs appearing under a smooth confinement potential have also attracted much theoretical attention. It was shown that this kind of ABS can be partially separated in
space and thus is dubbed a quasi-Majorana or partially separated ABS. Multiple theoretical works demonstrate that quasi-Majoranas with approximately opposite spins generate almost all the Majorana signatures except nonlocality and thus are indistinguishable from true MBSs in local measurements. It is worth noting that while quasi-Majoranas are not topologically protected due to the lack of nonlocality, they might be good enough to perform non-Abelian braiding using relative phase manipulations without spatial displacements \cite{vuik2019,sanJose2016,zeng2020}. 

\subsection{Distinguishing between Topological and Trivial}


\begin{figure*}[ht!]
    \centering
    \includegraphics[width=\linewidth]{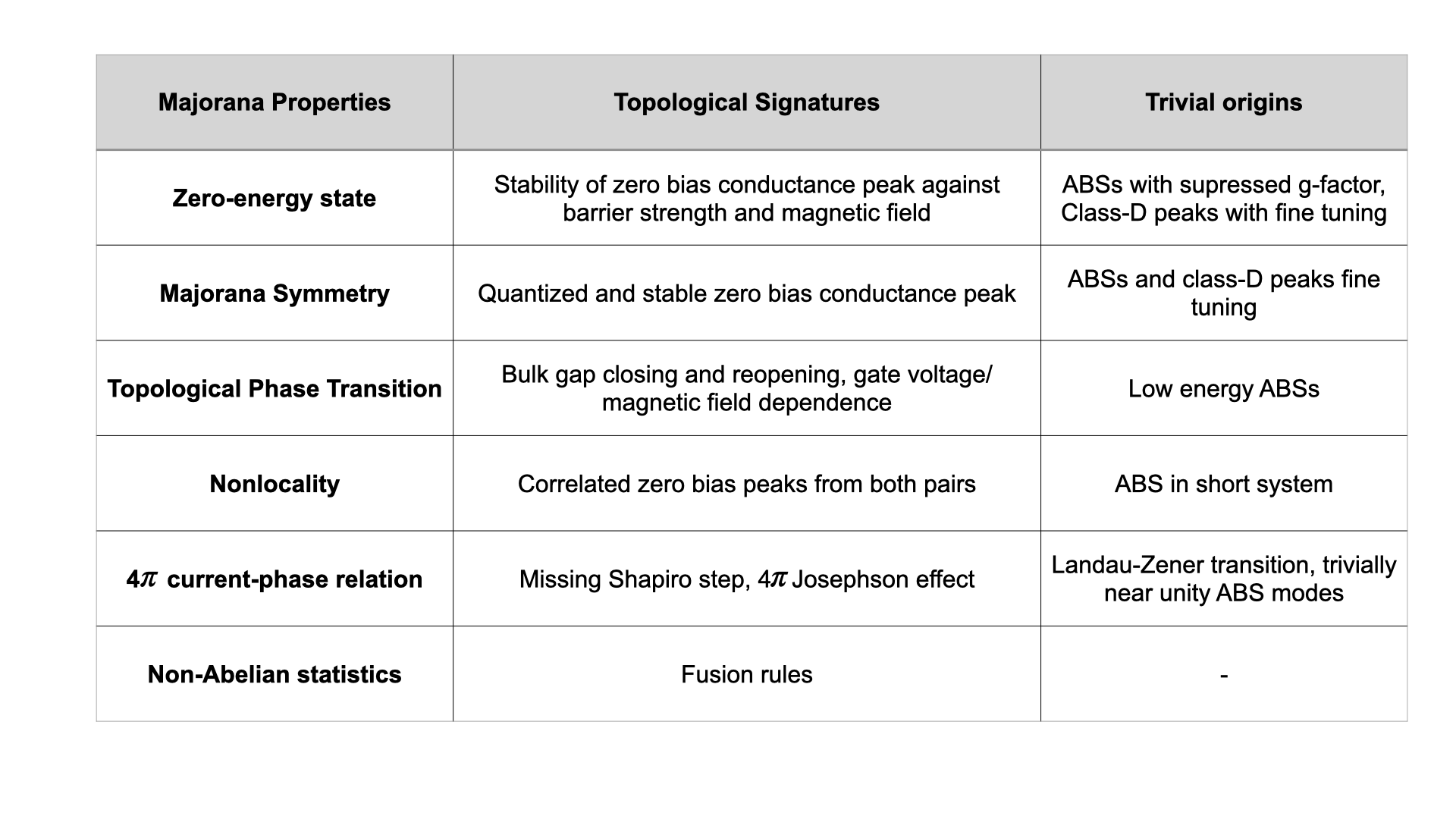}
    \caption{Classification of Majorana bound state properties and their corresponding experimental signatures together with their possible trivial explanations. }
    \label{fig:trivial vs topo}
\end{figure*}

The experimental search for MBSs in S--Sm hybrid devices has found many theoretically predicted signatures, especially the ZBCP in tunneling spectroscopy at finite magnetic fields\cite{mourik2012,das2012,deng2012,deng2016,albrecht2016,chen2017,fornieri2019,ren2019,banerjee2023_topoTransition}. Microwave \citep{rokhinson2012,laroche2019}, critical current \citep{ke2019}, and phase \citep{dartiailh2021_piJump} signatures have also been observed and considered similarly within the context of topological superconductivity.
However,
not all predicted signatures of MBSs have been observed, especially in a single device, such as
correlated and robust ZBCPs in an extended parameter space of gate voltage and magnetic field \cite{chen2019}, authentic quantized ZBCP \cite{yu2021},
the closing of the trivial bulk superconducting gap and the reopening of the topological gap \cite{huang2018_ABStoMBS}, and oscillations of the conductance peak with magnetic field \cite{dasSarma2012}. Additionally, experiments have reproduced many of the observed signatures in decidedly non-topological parameter regimes \citep{chen2019,yu2021,dartiailh2021_shapiro,elfeky2022} or invoked alternative explanations for their occurrence \citep{lee2012_kondoZBP,churchill2013,finck2013,zuo2017,valentini2021_fullShellRebuttal,haxell2023_orbital,kayyalha2020_nullMajorana}.
Theory has also suggested numerous alternative mechanisms to explain these features, including smooth inhomogeneities \citep{kells2012,roy2013,stanescu&tewari2014,cayao2015,fleckenstein2018,vuik2019,pan&dassarma2021_crossover,marra&nigro2022}
such as e.g.\ the presence of (intentional or accidental) quantum dots \citep{prada2012,liu2017_ABSvsMBS,setiawan2017,moore2018_MBSvsABS,moore2018_quantizedZBPs,pan2020_goodBadUglyZBCPs,pan2021_unquantized};
interband coupling \citep{woods2019_pinningMultiband};
Landau-Zener transitions \cite{dartiailh2021_shapiro};
and random disorder \citep{bagrets&altland2012,pikulin2012,liu2012,pan2020_goodBadUglyZBCPs,pan&dassarma2021_crossover,pan2021_unquantized,dassarma&pan2021,ahn2021,pan&dassarma2022_onDemandZBP,pan2022_zbpDisorder}.
Three papers were retracted \citep{zhang2018_retracted,*zhang2018_retractionNote,gazibegovic2017_retracted,*gazibegovic2017_retractionNote,he2017_retracted,*he2017_retractionNote}.



As the previous efforts were mainly focused on local tunneling spectroscopy from one end of the hybrid system, little has been studied about the bulk gap, the topological phase transition, and the nonlocality of the MBSs. Without knowing these properties, it will be difficult to distinguish an MBS from an ABS based solely on local spectroscopy \cite{pan2020_goodBadUglyZBCPs}. To solve this issue, new measurement methods have been explored. Among these, a three-terminal geometry that measures the two ends of the system simultaneously has been demonstrated to be a useful tool for distinguishing ABSs from MBSs \cite{yu2021,yu2023}. As MBSs should emerge simultaneously at the two ends of the topological region, three-terminal measurements that are capable of probing the two ends can utilize this unique property to help identify MBSs. In such three-terminal experiments, nearly quantized ZBCPs and accidentally correlated peaks from two ends have been observed \cite{yu2021}, and it was possible to rule out their topological origin by examining the local tunneling spectra from the two ends. Moreover, theoretical works proposed that the nonlocal conductance measured in the three-terminal geometry may detect the bulk gap closing and reopening as well as induced coherence length and thus offer new information about the hybrid system \cite{rosdahl2018,pikulin2021,pan2021_3terminal}. A possible drawback is that the nonlocal conductance may be small compared to local conductance and background noise. It is worth noting that a bulk gap closing and reopening signature may also arise due to the avoided crossing of ABSs in the bulk under the influence of a Zeeman field when the length of the hybrid region is comparable to the localization length of the induced superconductivity in the system \cite{hess2021}. Overall, simultaneous local measurements at the two ends combined with nonlocal measurements enabled by the three-terminal geometry shed light on the nature of conductance peaks in hybrid systems.

Going forward, a wealth of information can be imported from conventional superconducting qubits and circuit quantum electrodynamics.  These microwave techniques provide signatures of nontrivial topology that are complementary to d.c.\ transport \citep{mueller2013,ginossar&grosfeld2014,vaeyrynen2015,weston2015,dartiailh2017,harlechjones2022,pino2024} and enable rapid phase diagram mapping for device characterization \citep{razmadze2019,pikulin2021}.
Microwave spectroscopy also offers higher resolution than tunneling spectroscopy with better visibility into the low-lying or more crowded subgap spectral features in few-mode devices \citep{vanWoerkom2017,tosi2019,vanZanten2020,matuteCanadas2022,zellekens2022,sahu2024_parityFlipping,hinderling2024}.
Implementing microwave spectroscopy of planar junctions is complicated by the large number of conduction channels in high-density 2DEGs and typically requires reducing this number via gating \citep{chidambaram2022,hinderling2023_absSpec}.
We discuss microwave techniques and their utility further in \cref{sec:microwave}.

As summarized in \cref{fig:trivial vs topo}, while many predicted signatures have been observed in experiments, an unambiguous demonstration of MBSs is still inaccessible as most of them are accompanied by trivial explanations. Another fact that prevents the identification of MBSs is that most experiments only demonstrate one or two of the predicted signatures instead of a collection of signatures that corroborate each other. Given the similarities between MBSs and ABSs, more complex experiments that prove their nonlocality or topological protection are likely required to solve the controversy. Unequivocally establishing the existence of MBSs will likely require the direct observation of their non-Abelian statistics via fusion and braiding experiments and the development of a topological qubit.

\section{\label{sec:disorder}Disorder}

A number of theoretical studies suggest disorder present in current S--Sm materials needs to be further reduced to meet the requirements for topological superconductivity and MBS physics \cite{pan2022_zbpDisorder,pan2020_goodBadUglyZBCPs,yu2023}. However, few works have attempted to estimate the strength and origin of disorder from experiments in hybrid S--Sm systems.
It is quite challenging to separately identify different kinds of disorder as their observed signatures can be overlapping, and traditional semiconductor disorder signals in d.c.\ transport are obscured when the 2DEG is shunted by a high density metal or superconductor.

In our view, disorder remains the most intractable problem in hybrid systems, as structural disorders are incorporated during material synthesis and additionally introduced by material processing during nanofabrication.
We therefore focus this section on disorder: its various sources, its impacts on the topological phase, and ideas to reduce or mitigate its effects in future experiments.





\subsection{Disorder in Superconductors}

Proximity-induced superconductivity in the semiconductor is an indispensable requirement for topological superconductivity in hybrid systems. Variation of the order parameter in the superconductor or Cooper pair disorder could introduce disorder in the topological Hamiltonian.
Early works suggested that disorder in the superconductor is not as detrimental as impurities in the semiconductor or disorder at the interface because it is not pair-breaking \cite{stanescu2011,potter&lee2011_TIvsRashba_erratum} and does not suppress the topological superconductivity in the semiconductor \cite{lutchyn2012}. This is expected as s-wave superconductivity is robust to time-reversal invariant disorder \citep{anderson1959_dirty,abrikosov&gorkov1961_paramagnetic,abrikosov&gorkov1959_alloysZeroT} and hence proximity-induced superconductivity is robust to disorder in the parent superconducting bulk.
However, disorder in the superconductor could become non-negligible if the coupling to the semiconductor is strong.
Theory suggests that the strong-coupling regime is characterized by a proliferation of disorder-induced zero energy states and a reduced topological gap \citep{cole2016,awoga2019,awoga2022_disorder}.

We note that recent studies estimate the disorder in the superconductor on the scale of \SI{10}{meV}, which is much larger than previously reported and surprisingly even larger than the disorder in the semiconductor \cite{stanescu&dassarma2022}. These studies suggest that disorder in fact facilitates a large induced gap in the semiconductor by bridging the difference in Fermi energies at the S--Sm interface \cite{stanescu&dassarma2022}.

\subsection{Disorder in Semiconductors}

Disorder in semiconductors is mainly caused by crystal defects, which could in some form act as charge impurities. Theoretical works have studied the effect of charge impurities and demonstrated their adverse effects on the realization of MBSs \cite{ahn2021, woods2021}. Screening effects from the superconductor, metallic gate and free charge could generally reduce the magnitude and decay lengths of charge impurities. Screening is more effective when impurities are located near the S--Sm interface or the metallic gate\cite{woods2021}. 

While it is believed that disorder in semiconductors is a big obstacle to the realization of topological superconductivity \cite{pan2022_zbpDisorder}, experimentalists do not have the right measurements to identify the origin of disorder and less so to quantify the strength of disorder in hybrid devices. Recently, \citet{ahn2021} have estimated the disorder strength by fitting the transport mobility vs.\ density of experimental data into theoretical models. Such an indirect route has provided new insight about the order of magnitude of disorder strength in hybrid systems. Experimentally, disorder strength can be roughly estimated by analyzing the energy difference of delocalized states \cite{yu2023}. Overall, both theoretical and experimental studies indicate that the disorder strength is on the order of meV. While it is not clear how this disorder strength should be compared to stability of the topological superconductivity and role of screening, these numbers are at least an order of magnitude larger than the possible topological gap in current hybrid systems.

\begin{table*}[tbp]
\centering
\begin{threeparttable}
    \begin{tabular}{lS[table-format=2.2]SllcSSSSSl}
        \toprule
        {material} & {$\varepsilon/\varepsilon_0$} & {$m^*/m_\text{e}$} & {$n$ (\si{cm^{-2}})} & {$\mu_\text{tr}$ (\si{cm^2/Vs})} & {$\nu$} & {$B_\nu$ (T)} & {$V_\text{e-e}$ (K)} & {$B_\text{SdH}$ (T)} & {$\Gamma$ (K)} & {$\Gamma / V_\text{e-e}$} & {References} \\
        \midrule
        GaAs & 12.90 & 0.067 & 1.4e11 & 100e3 & 1/3 & 15.2 & 197 & 0.5 & 5.0 & 0.025 &  \\
        GaAs (H) & 12.90 & 0.2 & 9.4e10 & 200e3 & 2/3 & 5.8 & 122 & 0.3 & 1.0 & 0.008 &  \\
        AlAs & 10.06 & 1.1 & 2.1e11 & 77e3 & 2/3 & 13 & 233 & 1 & 0.6 & 0.003 &  \\
        ZnO & 7.8 & 0.234 & 3.9e11 & 45e3 & 4/3 & 12 & 289 & 0.5 & 1.4 & 0.005 &  \\
        graphene & 1{\tnote{a}} & 0 & 3.5e11 & 100e3 & 1/3 & 9 & 1954 & 0.25 & 13.8 & 0.007 & \citep{bolotin2008,bolotin2009} \\
        Si (111) & 11.6 & 0.188 & 4.15e11 & 325e3 & many, including 8/3 & 6.4 & 142 & 0.1 & 0.4 & 0.003 &  \\
        SiGe & 12.7 & 0.1505 & 2.7e11 & 60e3 & 2/3 & 16.8 & 210 & 0.35 & 1.6 & 0.007 &  \\
        InAs & 14.3 & 0.024 & 7.8e11 & 1.8e6 & 4/3 & 24 & 223 & 0.5 & 14.0 & 0.063 & \citep{ma2017} \\
        \bottomrule
    \end{tabular}
    \begin{tablenotes}
        \item[a] Suspended.
    \end{tablenotes}
\end{threeparttable}
\caption{\textbf{Electron-electron interactions and disorder in semiconductor electron and hole gases.}  We tabulate early observations of fractional quantum Hall states and Shubnikov-de Haas (SdH) oscillation onset fields in various material systems.  Magnetotransport features were observed at fractional filling factors $\nu$ at a magnetic field $B_\nu$.  The electron-electron interaction energy $V_\text{e-e} = (1/4\pi\epsilon)e^2/\ell_{B_\nu}$ is estimated from Coulomb's law at the magnetic length $\ell_{B_\nu} = \sqrt{\hbar/eB_\nu}$ representing the size/separation of the cyclotron orbits or ground state wavefunction.  The Dingle temperature $\Gamma = \hbar/2\tau$, representing the Landau level disorder-broadening, is estimated from the onset of SdH oscillations $B_\text{SdH}$ via the scattering rate $\tau = \omega_c^{-1}$ where $\omega_c = eB_\text{SdH}/m^*$ is the cyclotron frequency ($\omega_c = e.v_\text{F}B_\text{SdH} / \hbar(\pi n)^{1/2}$ for graphene, with $v_\text{F} \approx \SI{e6}{m/s}$).  Density $n$ and transport mobility $\mu_\text{tr}$ are also tabulated.  Dielectric constants $\varepsilon$ and carrier effective masses $m^*$ were obtained from Ref.~\onlinecite{adachi_semiconductors}.}
\label{tab:fqhe}
\end{table*}

While transport mobility is perhaps the most commonly cited metric for 2DEG quality, it may not be the best quantifier, especially when comparing devices across different material systems.  Aside from transport mobility, we should explore new methods of identifying disorder in semiconductor used for topological studies. For example the onset field of Shubnikov-de Haas oscillations $B_\text{SdH}$ obtained from quantum Hall data offers a unique metric.  This is an indicator of the disorder-broadening of Landau levels and is typically quantified by the Dingle temperature $\Gamma=\hbar\omega_c/2k_\text{B}$, where the cyclotron frequency $\omega_c$ at $B_\text{SdH}$ encodes the ability of a charge carrier to complete a cyclotron orbit in time $\omega_c^{-1}$ without scattering \citep{bolotin2008}. While transport is highly sensitive to back scattering, Dingle mobility provides an isotropic measure of scatterers. It is argued Dingle mobility is related to observation of fractional quantum Hall states where electron-electron interactions $V_\text{e-e}$ can manifest.  In \cref{tab:fqhe} we estimate the electron-electron interaction energy and Dingle temperature from early observations of the fractional quantum Hall effect in a variety of material systems.  We highlight that $\Gamma/V_\text{e-e}$ is relatively larger in InAs despite the much higher transport mobility due to the relatively smaller electron effective mass.  We also see that the disorder-broadening of the Landau levels in InAs as characterized by $\Gamma$ is roughly 2--10 times larger than in other materials where FQHE has been observed, with the exception of graphene. Graphene compensates for this broadening with strong electron--electron interactions, the interaction energy in InAs is much lower due to its relatively larger dielectric constant. This striking comparison shows that while buried InAs could have extreme high transport mobility, the Dingle mobility is quite limited in exploring fractional quantum Hall physics. Reducing disorder in these systems should focus on enhancing Dingle mobility in buried and surface quantum well structures.


A number of other known metrics could also be used to quantify the disorder and identify its various sources within the semiconductor heterostructure.  The quantum mobility associated with the onset of SdH oscillations further serves as a more stringent metric than transport mobility \citep{lodari2021}.  The metal-insulator transition, percolation density, and density-scaling of the conductivity are used to study how charged background impurities are screened and their origins \citep{shabani2014_metalInsulator,dassarma2005_metalInsulatorPercolation,dassarma&hwang2013,dassarma2013}.  By incorporating these and developing new metrics to better discriminate between different sources of disorder in the semiconductor heterostructure, materials scientists will be able to focus their efforts where they are needed the most.  A major outstanding challenge is to rigorously characterize the disorder in-situ in the topological devices of interest, as material processing during nanofabrication inevitably introduces additional sources of disorder.






\subsection{\label{sec:disorder:interface} Disorder at the Interfaces}

While individual characterization of the superconductor and semiconductor provide crucial design parameters, the eventual device is a hybrid combination of both materials. In addition, to achieve gate-voltage tunability of spin--orbit coupling and chemical potential, further material processing and deposition is required. Among the most important steps affecting device quality is dielectric deposition for the gate stack.

\subsubsection{\label{sec:s-sm-interface} Super/Semi Interface Quality}

As mentioned in \cref{sec:materials:gap}, a hard induced superconducting gap is a prerequisite for topological superconductivity in hybrid systems and is typically established via tunneling spectroscopy and quantized conductance doubling \citep{chang2015,kjaergaard2016,zhang2017_ballistic,gul2017,drachmann2017,sestoft2018,moehle2021,zhang2023_PbTe,valentini2024,chen2023_Pb-InSb,tosato2023}.  Interface homogeneity is crucial as disorder-induced subgap states will soften the induced gap and poison the topological superconducting state \citep{takei2013,potter&lee2011_TIvsRashba}.  Nevertheless, coupling the superconductor too strongly to the semiconductor can metallize the key semiconductor properties, like spin--orbit interaction and g-factor \citep{cole2015}, and transfer disorder from the superconductor, for example from disordered phases of silicides, to the semiconductor \citep{stanescu&dassarma2022}.  The thickness of the top barrier of the quantum well is therefore an important experimental knob for optimizing the coupling strength \citep{shabani2016, wickramasinghe2018}.


\begin{figure*}
    \includegraphics[width=\linewidth]{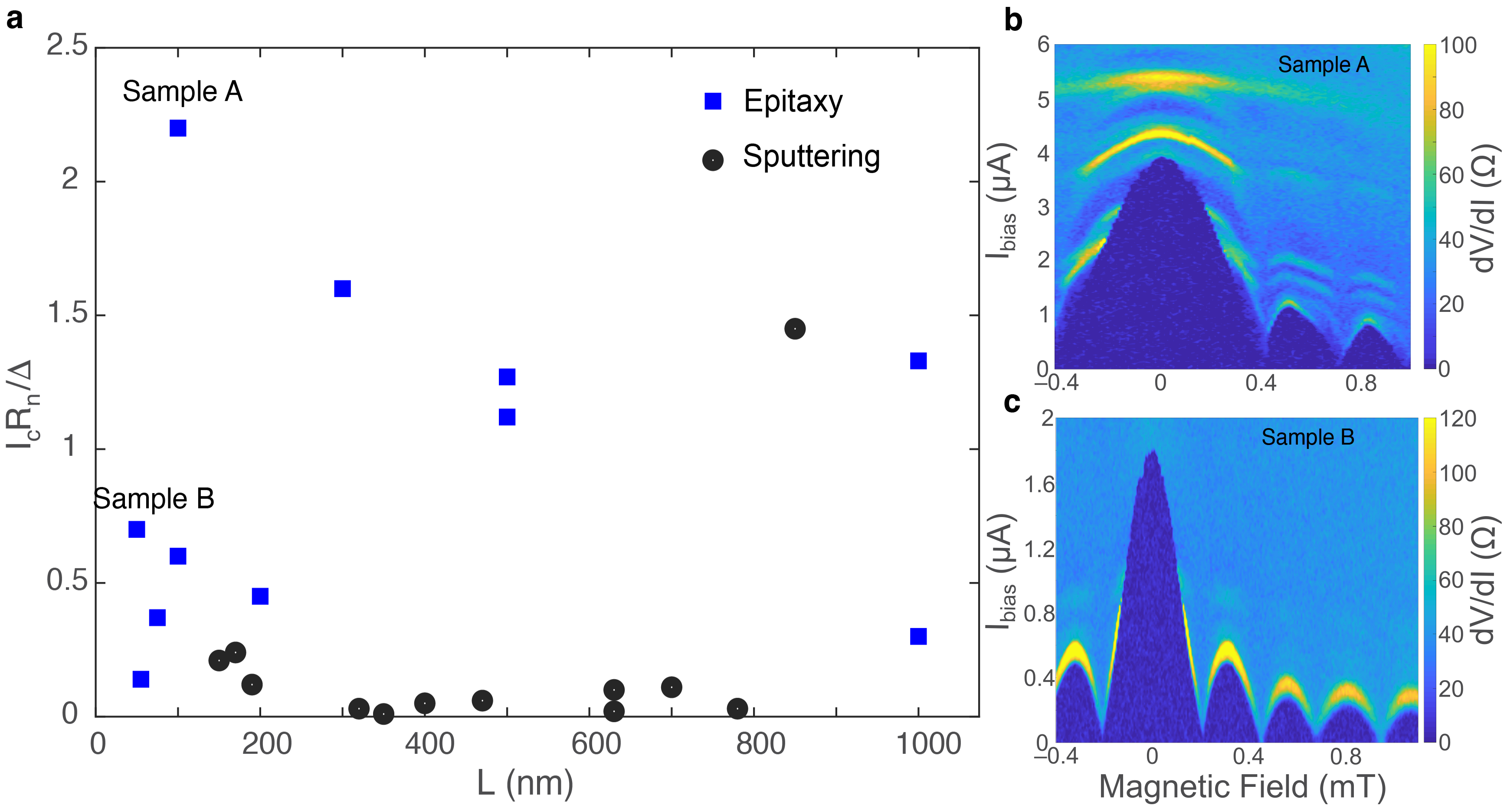}
    \caption[$I_\text{c}R_\text{n}$]{\textbf{$I_\text{c}R_\text{n}/\Delta$ as a metric for material and device quality.}  (a) Products $I_\text{c}R_\text{n}$ of critical current and normal resistance normalized by the superconducting gap $\Delta$ as a function of the distance $L$ between superconducting electrodes in a Josephson junction.  Junctions were formed either by epitaxial superconductor growth (blue squares) or by sputtering (black circles).  Data are compiled in Ref.~\onlinecite{yuan2021} Table I from Refs.~\onlinecite{mayer2019_proximitizedInAs,delfanazari_2017,Nitta92,TAKAYANAGI1995,Heida98,Zellekens2020,Perla2020,Gharavi_2017,mayer2020_proximitizedInAsSb,ke2019,wiedenmann_4-periodic_2016,Pallecchi2008} and Ref.~\onlinecite{mayer2019_proximitizedInAs} Fig.\ 4. (b,c) Differential resistance as a function of current bias and magnetic field threading the junction area for the data points labeled in (a).  Fraunhofer interference is observed along with resonances of different strength in the normal state.}
    \label{fig:IcRn}
\end{figure*}

The standard figure of merit for Josepshon junctions is the product $I_\text{c}R_\text{n}$ of its critical current $I_\text{c}$ and normal resistance $R_\text{n}$ \citep{tinkham,josephson1962,ambegaokar&baratoff1963,likharev1979review}.
The cancellation of geometric factors in the product allows for comparisons of different junctions of similar materials.
For example, the product can be related to the superconducting gap $\Delta$ in the zero-temperature limit as $I_\text{c}R_\text{n} = \chi \Delta/e$ where the constant prefactor $\chi$ depends on the relative length scales of the coherence length, mean free path, and weak link length delineating the ballistic/diffusive, short/long, and clean/dirty limits.
Thus the normalized product $I_\text{c}R_\text{n}/\Delta$ can be used to compare junctions formed from different superconductors.
However, a number of details should be considered when making $I_\text{c}R_\text{n}$ comparisons.
For instance, the experimentally observed switching current is a stochastic quantity that deviates from the theoretically defined critical current due to various effects, including phase and thermal fluctuations, quantum tunneling, and damping set by the junction resistance and capacitance.
Additionally, the relevant length scales in a hybrid system will generally depend on both the superconducting and semiconducting properties.
In an S--Sm--S junction, the normal-metal coherence length $\xi$ and mean free path $\ell$ of the proximitized semiconductor are compared along with the length $L$ of the semiconducting channel between the superconducting electrodes.
Al--InAs junctions are typically weakly in the short, ballistic and dirty regimes, with $L < \ell \lesssim \xi$.

\Cref{fig:IcRn}(a) shows a compilation of several Josephson junction studies of  nanowires and quantum wells where $I_\text{c}R_\text{n}/ \Delta$ is plotted vs $L$. The plot includes Al and Nb superconductors as well as InAs, InAs$_{0.50}$Sb$_{0.50}$, InSb, HgTe, and carbon nanotube semiconductors \cite{yuan2021}. Studies with in-situ epitaxial deposition of superconductors are plotted with blue squares while in-situ cleaning techniques and sputtering methods are plotted with black circles.  Epitaxial Josephson junctions outperform consistently in providing the highest $I_\text{c}R_\text{n}/ \Delta$ values as they involve little interface impurity physics. In addition there is a clear contrast in data taken from sample A with $I_\text{c}R_\text{n}/ \Delta > 1$ vs sample B with $I_\text{c}R_\text{n}/ \Delta < 1$, as shown in the Fraunhofer interference patterns of \cref{fig:IcRn}(b,c).  While both present satisfactory interference patterns, sample A shows many orders of presumably multiple Andreev reflections as bright peaks in differential resistance, indicating a well-ordered S--Sm interface.  Devices similar to sample A have shown supercurrent closing and reopening in the presence of an in-plane magnetic field \cite{dartiailh2021_piJump}, while in the authors' experience these signatures are not as clearly observed in devices with $I_\text{c}R_\text{n}/ \Delta \ll 1$ \cite{lotfizadeh2024}.





\subsubsection{Surface Dielectric for Gate Stack}
\begin{figure*}
    \centering
    \includegraphics[width=\linewidth]{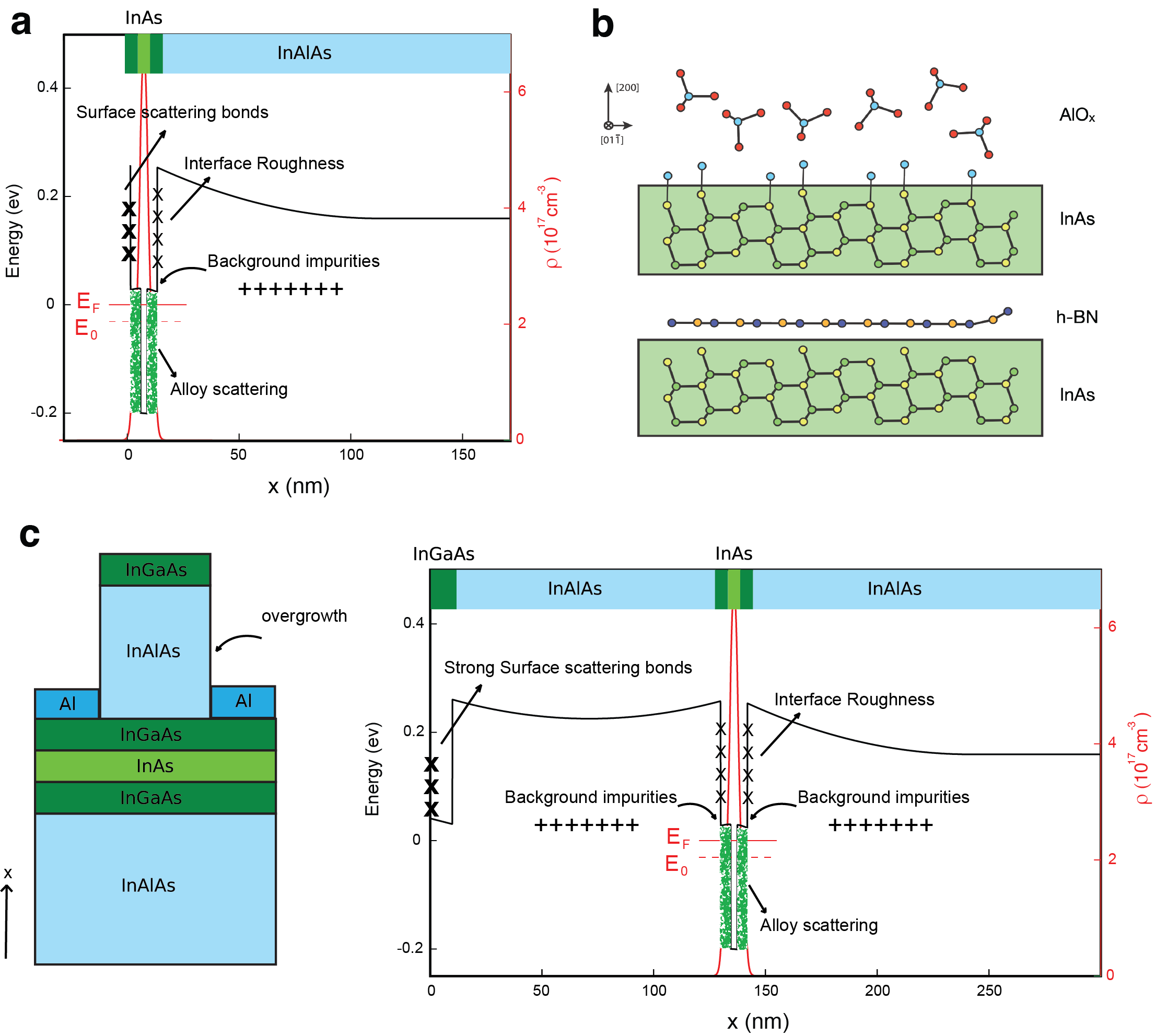}
    \caption[disorder in Sm and Sm-oxide interface]{\textbf{Disorder in the bulk and at the interfaces of the semiconductor heterostructure.}  (a) Electronic band structure (black) through the the depth $x$ of the heterostructure overlaid with the carrier density distribution $\rho$ (red).  Dangling bonds and interface roughness contribute to random scattering processes at the interfaces.  Remote ionized impurities in the bulk background dope the quantum well and increase carrier density.  Alloy scattering from the ternary barriers also disorder the 2DEG.  (b) Dangling bonds at the oxide-semiconductor interface form charge traps.  High-$\kappa$ dielectric 2D materials like hexagonal boron nitride (h-BN) can form a cleaner interface.  (c) In-situ overgrowth of insulating InAlAs removes the oxide interface and moves the surface farther away from the quantum well.  (d) Electronic band structure (black) through the the depth $x$ of the heterostructure overlaid with the carrier density distribution $\rho$ (red) in the overgrowth structure.  Background impurities on either side of the 2DEG dope the quantum well and act as remote scattering sites.  The strong surface scattering sites are shifted away from the quantum well and traded for weaker scattering from the interface roughness of both sides.}
    \label{fig:disorder}
\end{figure*}

Like most semiconductors, \RNum{3}-\RNum{5} semiconductors are easily oxidized when exposed to air and the oxide layer can eventually reach as thick as 10-20 nm \cite{tang1986} usually in amorphous phase. Oxidation process can increase surface roughness, and induce interface traps and defects, some in form of dangling bonds, which lead to an increased surface scattering \cite{pauka2020}. As a result, mobility in the semiconductor is further reduced severely \cite{pauka2020}. In early experiments where the superconductor needed to be deposited separately after nanowire growth, oxide on the nanowire surface would directly affect the transparency of the nanowire-superconductor interface. Over the years, methods such as argon cleaning, HF etching, sulfur passivation, and atomic hydrogen cleaning have been attempted to remove the oxide \cite{suyatin2007,gul2017,webb2015} before deposition of superconductors in an sputtering chamber with limited success. While these methods remove oxide more or less, it also damages the surface of semiconductor. It was the development of in-situ deposition of a superconductor that brought a truly oxide-free semiconductor-superconductor interface \cite{krogstrup2015, shabani2016}. The epitaxial 
Al grown in-situ leads to a significant improvement of the induced superconducting gap \cite{chang2015}. As epitaxial Al needs to be etched away to define the tunnel region in a nanowire device or to define the topological region for a planar Josephson junction, the oxide will still form on the surface of the semiconductor and cause reduced mobility. In a detailed study, Ref \cite{pauka2020} demonstrated it is possible to somewhat repair the degraded mobility by using argon-hydrogen plasma to clean the oxide followed by an in-situ protective ALD-grown $Al_2O_3$ coating. h-BN dielectric has also been tried recently as a non-chemical dielectric interface to InAs \cite{barati2021}.


A basic advantage of the gate-defined 1D wire geometry is that the topologically active region of the quantum well is directly below the superconductor, while in a planar junction the gate dielectric contacts the semiconductor in this region (\cref{fig:hybrids}(b,c)).  While the latter case allows for more direct gate tunability of the chemical potential and Rashba parameter in the well, it also introduces charge traps at the semiconductor-oxide interface that can act as scattering sites, as shown in \cref{fig:disorder}(a,b). As discussed, standard technique to add a gate stack is to introduce a dielectric, ALD-grown $Al_2O_3$ or $HfO_x$, or mechanical interfaces with h-BN. One fundamentally different way to alleviate the interface problem is to do an overgrowth. This involves an additional insulating III-V layer to be grown in-situ, as shown in \cref{fig:disorder}(c) after depositing the superconducting electrodes using an overgrowth technique after in-situ oxide desorption. This requires the fabricated devices to be reintroduced back to UHV MBE systems, the lattice matched dielectric will separate strong surface scattering with a buffer layer. \cref{fig:disorder}(c) shows this case specifically for InAs heterostructures. While transport mobilities in InAs surface quantum wells have been limited to 40,000-100,000 $cm^{2}$/Vs \cite{zhang2023_100kMobility,strickland2022_fermiLevelPinning,wickramasinghe2018} it is possible to achieve 500,000-1,000,000 $cm^{2}$/Vs \cite{hatke2017,shabani2014_gating} by burying the quantum well as proposed in Fig. \cref{fig:disorder}(c).  Importantly, as the superconductor-2DEG separation and the quantum well width remain unchanged, the proximity coupling is preserved and orbital effects are not increased.  Similar efforts involving in-situ junction fabrication have also been developed using shadow-wall lithography
\cite{mazur2022_diode,levajac2023_jjLength}
or selective area growth \citep{vaitiekenas2018_SAG}.

\section{\label{sec:devices} New Directions in Device Architectures}

Planar Josephson junctions must be long, with respect to the Majorana localization length, in the dimension transverse to the supercurrent to spatially separate MBSs. Beyond this geometrical constraint, however, the simplistic theoretical picture of \cref{sec:theory} does not provide more insight about the shape of the superconducting lead. There has been great progress in recent years in identifying various geometries of superconducting leads to enhance the topological transition parameters and conditions with the same materials stack. These developments are generic and can be applied to future proximitized materials systems.

\subsection{\label{sec:Wsc}Superconducting Contact Width}

\begin{figure}
    \centering
    \includegraphics[width=\linewidth]{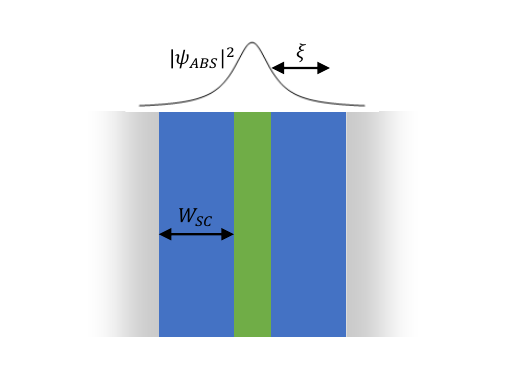}
    \caption[Superconducting contact width]{Effect of superconducting contact width $W_\text{SC}$ on the
        character of Andreev bound states (ABSs) in superconductor--semiconductor--superconductor junctions.
        ABSs are localized in the semiconducting junction region (green)
        but decay into the superconducting contacts (blue) over a characteristic length
        scale given by the coherence length $\xi$ of the induced superconductivity.
        Depending on the ratio $W_\text{SC}/\xi$, the ABS wavefunction
        $\psi_\text{ABS}$ depends on the degree of normal reflection at the blue-grey
        interface behind the contacts.
        The grey region may be characterized by e.g.\ a metallic or insulating density of
    states.}%
    \label{fig:Wsc}
\end{figure}

Geometric effects on the topological phase in planar junctions are typically modeled by spatial variations of the parameters in \cref{eq:2D-p+ip-hamiltonian,eq:1D-majorana-wire-hamiltonian}.
For example, the superconducting order parameter is commonly taken to be nonvanishing only beneath the superconducting contacts (see \cref{fig:Wsc})
\begin{equation}
    \Delta(y) = \begin{cases}
        0 & \text{if } \abs{y} < W_\text{Sm}/2 \\
        \Delta e^{i\text{sign}(y)\varphi/2} & \text{if } W_\text{Sm}/2 < \abs{y} <
        W_\text{Sm}/2
        + W_\text{SC}.
    \end{cases}
    \label{eq:order-parameter}
\end{equation}
Both infinite \citep{pientka2017} and finite \citep{hell2017_planarJJ,setiawan2019_narrowing,sharma&mohanta2024} systems have been studied theoretically.
Experiments tend to categorically fall within three regimes defined by the (induced) superconducting coherence length $\xi$:
\begin{enumerate*}[label=(\alph*)]
    \item $W_\text{SC} \ll \xi$ \citep{fornieri2019}, where the topology of the system is highly sensitive to the chemical potential $\mu$ but insensitive to the superconducting phase drop $\varphi$;
    \item $W_\text{SC} \gg \xi$ \citep{ren2019}, where the topology is sensitive to $\varphi$ but not $\mu$; and
    \item $W_\text{SC} \sim \xi$ \citep{dartiailh2021_piJump}, where the topology is relatively sensitive to both.
\end{enumerate*}
$W_\text{SC}$ is therefore an important device parameter that determines whether the system topology is more gate-tunable or flux-tunable.
A systematic study of $W_\text{SC}$ and its effect on the anomalous Josephson phase and critical current minima with respect to Zeeman field---both relevant to topological signatures---has only recently been carried out in the regime $W_\text{SC} \lesssim \xi$ \citep{haxell2023_orbital} and emphasizes the importance of orbital effects in the planar junction geometry with external magnetic field \citep{pientka2017}.
Few systematic studies of the junction gap width $W_\text{Sm}$ have also been carried out \cite{ke2019,levajac2023_jjLength}, despite the important role it plays in determining the critical field at which the induced gap closes and whether the junction is wholly diffusive or partly ballistic.
\begin{figure}[ht!]
  \centering
  \includegraphics[width=\linewidth]{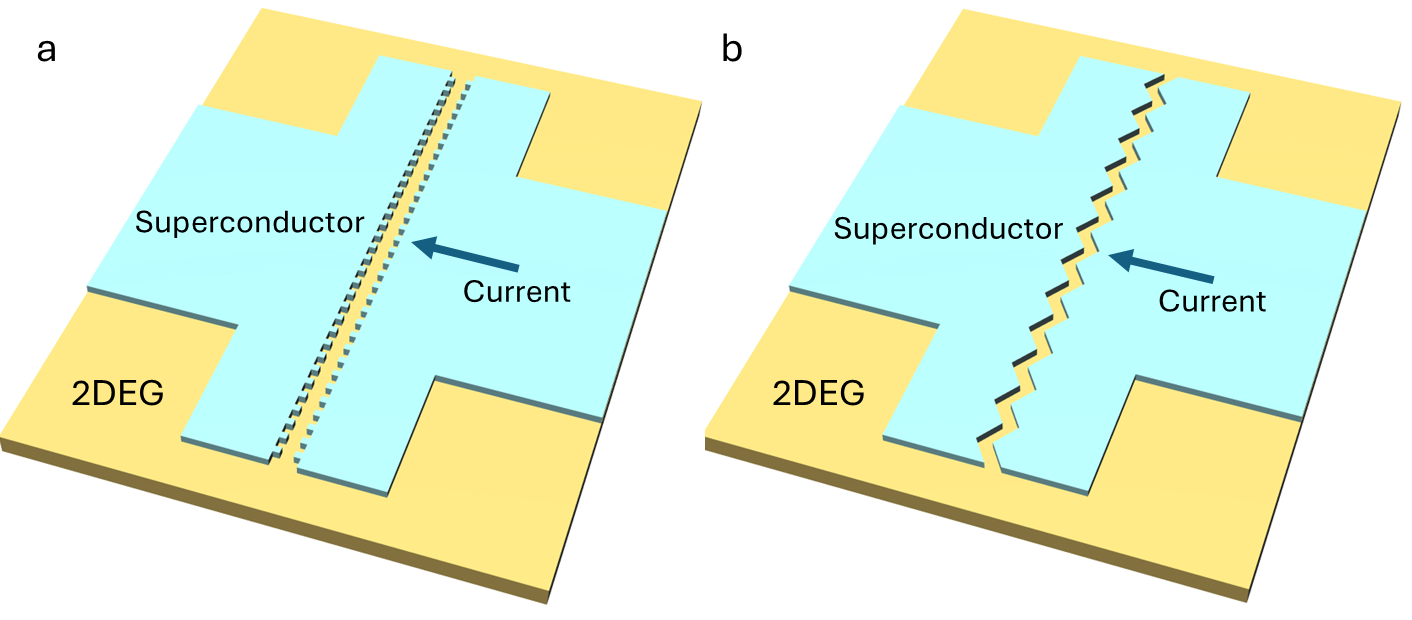}
  \caption{Illustration of (a) a junction with periodically modulated width and (b) a zigzag-shaped junction. Both geometries are supposed to enhance the topological gap \citep{laeven2020,paudel2021}.
 }
 \label{fig:Junction geometries}
\end{figure}

\subsection{\label{sec:periodic}Junction Gap Geometry -- Zigzag and Periodic Structures}

Optimization of the geometry of the quasi-1D channel is also an opportunity in planar junctions and gate-defined wires.
Periodic structures promise to increase the topological gap \citep{woods&stanescu2020,paudel2021,laeven2020} and enlarge the topologically nontrivial phase space volume \citep{adagideli2014,woods&stanescu2020,paudel2021,schirmer2024} while relaxing constraints on the chemical potential \citep{lu2016_superlattice,levine2017,escribano2019} and expanding the operational range of carrier densities in topological junctions.

The electrons with large characteristic wave vectors in the low density regime are less likely to be localized by disorder \cite{yu2023}; however, the chemical potential is largely set by the composition of the material heterostructure and band bending effects at its interfaces. Unfortunately, the topological condition \eqref{eq:1D-topological-condition} constrains the chemical potential in topological junctions to the low-density regime, where screening of the disorder potential due to random charged impurities is reduced. However the application of a strong periodic potential within the junction channel can lift this constraint by producing a miniband structure, with higher-energy minibands characterized by an increased effective Rashba spin--orbit coupling and therefore a larger topological gap \cite{paudel2021}. 
The increased effective Rashba spin--orbit coupling, topological gap, and characteristic wave vectors concomitant with a higher chemical potential all contribute to a topological phase that is more robust against disorder. Such a periodic potential can be realized by spatially modulating the width of the planar Josephson junction as illustrated in \cref{fig:Junction geometries}(a) \cite{paudel2021} or by making perforated superconducting contacts \cite{banerjee2023_topoTransition}. 

Engineering a periodic zigzag-shaped junction gap is also discussed as a way to eliminate the junction modes characterized by long semiclassical trajectories propagating at oblique angles to the superconducting contacts \citep{laeven2020} as shown in \cref{fig:Junction geometries}(b).
These so-called long-junction modes have a small Thouless energy due to their long dwell time within the weak link. They appear at high densities with subgap bound-state energies disconnected from the continuum and therefore contribute to a soft induced gap and small topological gap. By truncating these modes geometrically, high-density topological regimes with enhanced topological gaps are accessible.
Experimentally, long-junction trajectories are scattered out by disorder in the weak link, setting an upper limit to the trajectory length on the order of the semiconductor mean free path. As material improvements continue to increase the mobility of the semiconducting 2DEG, we expect these types of geometrical optimizations to become increasingly relevant and advantageous.

\subsection{\label{sec:phase}Alternative Sources of T-symmetry Breaking}

In 1D and 2D S--Sm devices, a Zeeman interaction realized with an applied magnetic field is commonly used to break time-reversal symmetry. While the application of a suitable Zeeman field is not difficult, it weakens the proximity-induced superconductivity in the semiconductor, which often leads to a soft gap and ubiquitous subgap states. An advantage of the planar junction geometry over 1D wires is that the requirement of a Zeeman field can be partially alleviated by a phase bias. Specifically, when the superconducting phase difference across the junction is $\pi$, the topological phase emerges close to zero Zeeman field \cite{hell2017_planarJJ,pientka2017}. However, the topological gap remains small in the absence of a strong Zeeman field.  Arrays of Josephson junctions could offer flexibility and a high degree of control over the phases and topological gap at the expense of a more complex device architecture \citep{lesser2024_4phase}.

With more complicated device designs, the requirement of a Zeeman field is further reduced or even eliminated by utilizing a phase winding or supercurrents counterpropagating along the junction as the source of time-reversal symmetry breaking\cite{melo2019,lesser2021_3phase,lesser2021_phaseInduced,lesser2022}. Flexibility in device geometry is a great advantage of the planar junction platform over the 1D wire system. With optimized device and junction geometries, future devices may have enhanced properties while requiring less demanding materials and fewer external controls.



\subsection{\label{sec:microwave}Microwave integration}

\begin{figure*}
    \centering
    \includegraphics[width=\linewidth]{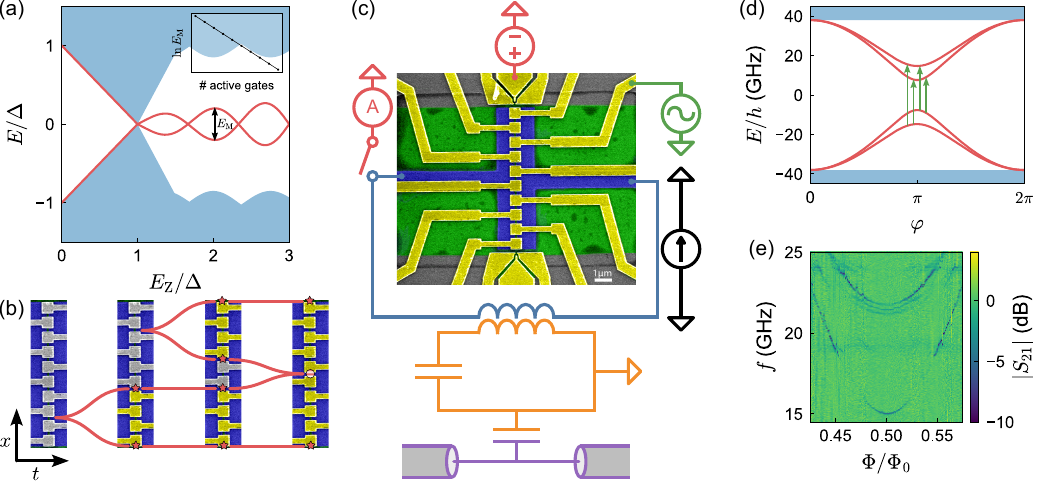}
    \caption{\textbf{Microwave integration, spectroscopy, and fusion in a planar Josephson junction.}  \textbf{(a)} Schematic spectrum showing Majorana oscillations as a function of Zeeman energy $E_\text{Z}$, in units of the induced gap $\Delta$.  The quasiparticle continuum (blue) and lowest-energy states (red) are shown.  The inset depicts the exponential dependence of the amplitude $E_\text{M} \propto \exp(-L/\xi_\text{M})$ on the length $L$ of the active region, controlled by an array of gates, in units of the Majorana localization length $\xi_\text{M}$.  \textbf{(b)} Non-trivial fusion protocol of Majorana bound states (MBSs) (stars).  Red lines are MBS worldlines.  Active (inactive) gates are yellow (gray).  \textbf{(c)} False-colored scanning electron micrograph of a superconductor(blue)--semiconductor(green) planar junction with gates (yellow).  The junction is embedded in a d.c.\ tunneling spectroscopy circuit (red) and a microwave spectroscopy circuit comprising a SQUID (blue) inductively coupled to a readout resonator (orange) capacitively coupled to a transmission line (purple).  An r.f.\ drive (green) is applied to a gate, and a flux line (black) controls the junction phase $\varphi$ via the flux $\Phi$ through the SQUID loop.  \textbf{(d)}  Andreev bound state (ABS) spectrum $E(\varphi)$ as a function of junction phase $\varphi$ depicting two pairs of modes of different transparency (red) and the transitions between them (green arrows).  \textbf{(e)} Measured ABS transition frequencies in terms of transmission $S_{21}$ as a function of flux $\Phi$ and drive frequency $f$.}
    \label{fig:microwave}
\end{figure*}

While the search for MBSs in S--Sm devices initially relied heavily on d.c.\ transport measurements like tunneling spectroscopy, attention has gradually turned to microwave techniques.  This progression is natural, as operations like braiding and fusion required for demonstrations of non-Abelian statistics and eventual qubit operation should be carried out in the microwave domain \citep{contamin2021,schmidt2013,aasen2016}.  Though the dominant decoherence mechanisms of the hypothetical Majorana qubit have yet to be established empirically,
parity-flipping processes like quasiparticle poisoning or thermal excitations set a lower bound on its operational frequencies.
An upper frequency bound is set by the requirement of adiabaticity to avoid driving excitations out of the topologically protected subspace of the ground state manifold, with a topological gap of $\SI{20}{\micro eV}$ setting an upper limit of \SI{5}{GHz}.  Radio frequency reflectometry or transmission measurements provide fast single-shot parity-to-charge readout of the fusion results \citep{vigneau2023review,wesdorp2023,razmadze2019,jong2019,karzig2017}.

En route to a topological qubit, microwave techniques provide valuable information about S--Sm devices and their constituent materials.  Studies of quasiparticle dynamics \citep{hays2018,bargerbos2022_singletDoubletQD,elfeky2023,erlandsson2023} 
provide insight into decoherence mechanisms like quasiparticle poisoning, while spectroscopic measurements reveal signatures of spin--orbit coupling and Coulomb interactions from which their strengths can be estimated in-situ \citep{vanWoerkom2017,tosi2019,matuteCanadas2022,bargerbos2023_andreevSpec} along with the carrier density and mobility of the proximitized semiconductor \citep{phan2022_pairing}.  In \cref{fig:microwave} we show a measurement setup integrating both d.c.\ transport and microwave spectroscopy.  The device (\cref{fig:microwave}(c)) is a planar junction with an array of independently tunable gates.  A second layer of gates (not shown) can be used to control the spaces in the first layer.  The gates can be used to form quantum point contacts \citep{chidambaram2022,hinderling2023_absSpec} in the topologically trivial regime at various positions along the junction to obtain both spatial and spectral information in the microwave domain, as shown in \cref{fig:microwave}(d,e).  Additional quantum point contacts at the junction edges enable tunneling spectroscopy to probe for spectral signatures of MBSs such as Majorana oscillations (\cref{fig:microwave}(a)), from which the Majorana localization length can be inferred by varying their separation via the junction gates \citep{prada2020review,albrecht2016}.  Sufficiently localized MBSs should exhibit nontrivial fusion rules, as illustrated in \cref{fig:microwave}(b) \citep{souto&leijnse2022,zhou2022_vjjGate}.



\subsection{Braiding and Fusion}

The promise of topological quantum computation relies on braiding transformations to implement the desired computational gates. These are operations on the degenerate ground state, e.g. moving MBSs adiabatically with respect to the topological gap, which result in entanglement and quantum gates. In Abelian phases, these movements only result in a trivial phase factor and there is no path for superposition and entanglement. The non-Abelian exchange statistics of MBSs enable them to perform braiding operations, in which exchanging the positions of two MBSs leads to a non-trivial rotation in the ground state manifold. Non-Abelian braiding together with topological protection is thus the foundation of topological quantum computation. Moreover, a demonstration of non-Abelian braiding is unequivocal proof of the existence of MBS and their topological nature. Bearing so much importance, braiding and fusing MBSs in a controlled manner is, therefore, the ultimate prize of Majorana research and will open a path toward topologically protected, fault-tolerant, quantum computation. It is worth noting that braiding Majoranas is not enough for universal quantum computing due to its incapability to generate arbitrary single-qubit rotations and two-qubit gate operations \cite{aguado2017review}. One possible solution to solve this issue is to use other quantum gates, which are not topologically protected, to perform these two operations along with braiding. Alternatively, one may also couple other standard qubits to Majorana qubits to complete the requirements for universal quantum computing \cite{hassler2010,leijnse&flensberg2011}. One should also stress that the Gottesman-Knill theorem \citep{gottesman1998_gkTheorem} renders Majorana braiding operations classically simulable in polynomial time \citep{calzona2020}. A full discussion of braiding and its advantages and limitations can be found in Refs.~\citenum{nayak2008review,dasSarma2015review}.
 
Any braiding and fusion operation has the prerequisite of creation, manipulation, and measurements of Majoranas. While braiding experiments may still be in future it is still useful to briefly introduce some theoretical proposals that realize braiding since they require different device architectures.  
 
A straightforward way to perform braiding is to physically move MBSs in a network of 1D systems. By using local gates \cite{chen2017, elfeky2021_minigate}, one may exchange the positions of two Majoranas in T-shaped nanowire or Josephson junction networks by controlling the tunnel coupling \cite{alicea2011,sau2011}. One difficulty of using local gates in hybrid systems is the screening from superconductors, which is worse in nanowire systems. Other proposals also suggest using Coulomb interactions \cite{vanHeck2012,hyart2013}, supercurrent \cite{romito2012} or phase differences \cite{liu2019_fluxInduced,zhou2020_xjjPhase} as the control knob for braiding MBSs. The Coulomb interaction is controlled by, for example, the magnetic flux in a Josephson junction array \cite{vanHeck2012,hyart2013}. Measurement-only braiding scheme based on a series of topological charge measurements has also been proposed \cite{bonderson2008,plugge2017}. The measurement-only scheme eliminates the need to physically move the MBSs but relies on the realization of parity-dependent interferometry \citep{dasSarma2015review}. Recent works based on that scheme also demonstrate the advantage of protecting the system from quasiparticle poisoning \cite{karzig2017}. A major challenge is the measurement of the parity operator of MBSs for two pairs and above (more than 4 MBSs). 

While topologically protected braiding demands fully nonlocal MBSs, non-protected braiding may merely need a local probe being selectively coupled to a single MBS \cite{prada2017}. Since quasi-Majorana states can have exponentially different coupling across a tunnel barrier, braiding is supposed to work with topologically trivial quasi-Majorana states using, for example, Coulomb-assisted braiding scheme \cite{vuik2019}. Such parametric non-Abelian braiding is dubbed as poor man's braiding due to the lack of topological protection. 
Generally, braiding of quasi-Majorana states could be an attractive approach
to demonstrate braiding properties given that the errors introduced by the measurement processes are carefully controlled \citep{zeng2020}.






\section{Outlook}

In this work, we gave our perspective on the progress and prospects of
realizing topological superconductivity and MBSs.
We highlighted the key advantages and challenges for different materials and devices, with a focus on epitaxial S--Sm materials and planar Josephson junctions.
The numerous requirements call for materials selection and growth informed by nanofabrication and  device engineering.
Currently, epitaxial Al--InAs is the most explored heterostructure for studying topological superconductivity in both quantum wires and planar Josephson junctions.

Conclusive observation of MBSs remains elusive due to the myriad of trivial mechanisms yielding similar signatures.
Part of the challenge is due to the presence of ABSs and quasi-MBSs and the limitations of current detection methods.
Given these complexities, it is important to explore new methods that could help to distinguish MBSs from their topologically trivial counterparts.
For example, microwave spectroscopy and other r.f.~techniques are natural successors to tunneling spectroscopy and d.c.~transport, and we believe the recent progress on that front will continue to bear fruit.

We discussed the origins and effects of disorder in Al--InAs and how it can be improved.
These guidelines are generic for quantum well structures.
Enhancing the topological gap by increasing the parent gap should be further explored hand in hand with the development of high spin--orbit materials.
In addition, clever device geometries can be designed to enhance the topological gap.
Designs involving phase biasing and supercurrent as alternative sources of time-reversal symmetry breaking can mitigate the requirement of a Zeeman field but further experimental studies are needed.  

We conclude by stating that although unequivocal observation of MBSs is still a work in progress, systematic and intertwined studies of quantum materials and devices are likely to yield advances in pursuit of that goal. In the effort to unveil the nature of MBSs, the past decade has seen major advances in materials synthesis and device fabrication, and has markedly expanded our understanding of mesoscopic superconductivity and Andreev physics.

\section{Acknowledgements}

We acknowledge support from DARPA TEE award no.\ DP18AP900007 and ONR MURI award no.\ N00014-22-1-2764. W.F.S.\ acknowledges support from the NDSEG Fellowship.  We also thank members of the Center for Quantum Information Physics at NYU for providing feedback and comments on this manuscript.  J.S.\ likes to thank Rosa Alejandra Lukaszew for inspiring discussions over the years on this topic. 

\bibliography{bibs/PRXQ_MERGED_AND_DEDUPED}
\end{document}